\documentclass[12pt,epsf,amssymb,qsymbols,axodraw]{article}
\usepackage{tabularx}
\usepackage{array}
\usepackage{graphics}
\usepackage{graphicx}
\usepackage{psfrag}
\usepackage{epsfig}
\usepackage{amsmath}
\usepackage{amssymb}
\usepackage{color}
\usepackage{rotating}
\makeatletter

\usepackage{verbatim}
\newcommand{\msbar}{{\overline{\rm MS}}}

\newcommand{\bea}{\begin{eqnarray}}
\newcommand{\eea}{\end{eqnarray}}
\newcommand{\be}{\begin{equation}}
\newcommand{\ee}{\end{equation}}
\newcommand{\gev}{{\rm GeV}}

\newcommand{\ra}{\rightarrow}

\newcommand{\pdir}{p\kern -5.2pt\raise 0.2ex\hbox {/}}
\newcommand{\vdir}{v\kern -5.75pt\raise 0.15ex\hbox {/}}
\newcommand{\kdir}{k\kern -5.75pt\raise 0.15ex\hbox {/}}
\newcommand{\epsdir}{\epsilon\kern -5.0pt\raise 0.15ex\hbox {/}}
\newcommand{\bvdir}{\bar{v}\kern -5.75pt\raise 0.15ex\hbox {/}}
\newcommand{\Ddir}{D\kern -7.75pt\raise 0.20ex\hbox {/}}
\newcommand{\Adir}{A\kern -7.75pt\raise 0.20ex\hbox {/}}
\newcommand{\ldir}{l\kern -5.0pt\raise 0.2ex\hbox{/}}
\newcommand{\varepsdir}{\varepsilon\kern -5.5pt\raise 0.15ex\hbox{/}}

\newcommand{\nn}{\nonumber}

\def\lsim{\raise0.3ex\hbox{$<$\kern-0.75em\raise-1.1ex\hbox{$\sim$}}}
\def\gsim{\raise0.3ex\hbox{$>$\kern-0.75em\raise-1.1ex\hbox{$\sim$}}}
\def\noi{\noindent} \def\nn{\nonumber} \def\bea{\begin{eqnarray}}
\def\eea{\end{eqnarray}} \def\beq{\begin{equation}}
\def\eeq{\end{equation}} 
\makeatother

\begin{document}

\thispagestyle{empty} 
\vspace*{17mm}

\begin{center}
{\Large \bf Spatial distributions in static heavy-light mesons:}\\
\vspace{.2cm}
{\Large \bf  a comparison of quark models with lattice QCD}\\
\vskip 1.25cm\par
{\scalebox{.92}{\par\centering \large  
\sc D.~Be\'cirevi\'c$^{a}$, E.~Chang$^b$, }}
{\scalebox{.92}{\par\centering \large  
\sc L.~Oliver$^a$, J-C.~Raynal$^a$, A.~Le~Yaouanc$^a$}}
{\par\centering \vskip 0.35 cm\par}
{\sl 
$^a$Laboratoire de Physique Th\'eorique (B\^at.~210)~\footnote{Laboratoire de Physique Th\'eorique est une unit\'e mixte de recherche du CNRS, UMR 8627.}\\
Universit\'e Paris Sud, Centre d'Orsay, F-91405 Orsay-Cedex, France.\\
$^b$Dept.d'Estructura i Constituents de la Mat\`eria, Institut de Ci\`encies del Cosmos (ICC), 
Universitat de Barcelona, Mart\`i Franqu\`es 1, E08028-Spain. }

\end{center}

\vskip 0.75 cm
\begin{abstract}
Lattice measurements of spatial  distributions of the light quark bilinear densities in static mesons allow one to test directly and in detail the wave functions of quark models. These distributions are 
gauge invariant quantities directly related to the spatial distribution of wave functions. We make a detailed comparison of the recent lattice QCD results with our own relativistic quark models, formulated
previously for quite different purposes. 

We find a striking agreement not only between our two quark models, but also with the lattice QCD data for the ground state in an important range of distances $r\lesssim 4~$GeV$^{-1}$. 
Moreover the agreement extends to  the $L=1$ states [$j^P=(1/2)^+$]. An explanation of several particular features completely at odds with the nonrelativistic approximation is provided. 
A rather direct, somewhat unexpected and of course approximate relation between wave functions of certain quark models and QCD has been established. 
\end{abstract}

\renewcommand{\thefootnote}{\arabic{footnote}}
\newpage
\setcounter{footnote}{0}
\section{Introduction}

The impressive phenomenological success of the quark model can be most directly apprehended in the spectroscopy of hadron states, where a huge number of  predicted  states is actually observed. A well-known example is the rich spectrum of baryons. 
However, the success of  quark models is often unfairly restricted to their success in spectroscopy, while they  successfully predicted many transition matrix elements between hadrons. In some cases this basically reflects SU(6) symmetry, like for the ratios of magnetic moments. In other cases, instead, the success relies on a detailed knowledge of the wave functions:  a striking example is that of the Okubo-Zweig-Iizuka allowed decays of radially excited states of charmonium  which depend on the precise positions of the nodes of radial wave functions.

Even though the quark model wave functions cannot be related to quantum field theory in a rigorous way, their phenomenological success requires a more detailed study and comparison with quantities that can be computed in field theory. 
It is our aim to show that spatial distributions of bilinear densities, which in quark models are very directly connected with wave functions, are indeed in agreement with results obtained in lattice QCD when one of the valence quarks is static. For certain models discussed below, we will show that this agreement holds true for a wide range of distances.  In this way we directly demonstrate the relevance of the wave function picture. 

Our study is not the first one in the field. To our knowledge, the first attempts in this direction were made in  ref.~\cite{duncan} where the authors predicted the form of equal-time Bethe-Salpeter amplitudes for heavy-light mesons in the Coulomb gauge. Their findings were in good agreement with wave functions deduced from a certain relativistic quark model.  Although encouraging, the problem with these results was that the quark models do not refer to any specific gauge, in contrast to the above lattice calculations. It is true that the Coulomb gauge, as  used in ref.~\cite{duncan}, is often considered as more physical, yet it is not compelling to identify it with the quark model treatment.

Another lattice determination of Bethe-Salpeter amplitudes was reported in ref.~\cite{michael}, where a qualitative agreement  with quark model expectations was shown for various meson states (see their fig.~5 in ref.~\cite{michael}). Even though they use a Wilson line to preserve gauge invariance of the  Bethe-Salpeter amplitudes, the relation with quark model wave functions is not straightforward.

A very useful contribution has been provided by the UKQCD-Helsinki group in ref.~\cite{koponen}. The idea was to consider a light quark in the field of a static heavy quark, and consider  spatial distributions of the charge and the matter densities, both computed on the lattice~\footnote{In fact, such distributions were already considered by UKQCD in ref.~\cite{dedivitiis} in the computation of the axial coupling $g_{B^\ast B\pi}$ in the static heavy quark limit. See also a discussion made in ref.~\cite{di-pierro0104208}.} and then compared with corresponding predictions made within a certain Dirac-type quark model.
\noindent There are several advantages in considering spatial distributions :
\begin{itemize}
\item[(1)] This directly eliminates the problem of reference to a gauge, since they are gauge invariant quantities: one simply calculates the matrix elements of the light quark current at a distance $\vert \vec{r}\vert$ from the static quark.
\item[(2)]  The same spatial distributions are easily calculable in quark models, and a rather direct conceptual connection is maintained with the quark model wave functions. This connection is especially direct in the case of Dirac-type quark models, since these spatial distributions are then simply quadratic combinations of the upper and lower components of the Dirac wave functions. This general type of model has precisely been considered by the UKQCD-Helsinki group and compared to their measurements of charge and matter (vector and scalar) density distributions. Other types of models, like the Bakamjian--Thomas approach, also yield these quantities rather straightforwardly, as we show in the present paper.
\item[(3)] Moreover a connection with physical observables is more direct. A simple integration over these densities yields quantities of direct physical interest, such as the pionic decay couplings of $1^- \ra 0^-$ (e.g. $B^\ast \to B\pi$), and  $1^+ \ra 0^+$ (e.g. $B_1^\prime \to B_0^\ast \pi$) often referred to as $\widehat g$ and $\widetilde g$~\cite{damir0406296}.
\end{itemize}
We find a few difficulties in the work of ref.~\cite{koponen}. In their lattice calculation, they claim to determine an extensive spectrum of static heavy-light mesons, including the higher orbital and radial excitations, up to $L=3$ and $n=3$, with surprisingly massive radially excited states. When trying to reproduce these results by   quark models, they are led to consider rather uncommon confining potentials that are a mixture of scalar and vector parts with a very large string tension. With such potentials, they do not reproduce the charge distribution within the doublet of lightest mesons, $j^P=(1/2)^-$, which should in principle be the one for which the models work best.

The aim of the present work is to revisit that problem by using more standard quark models, based on a scalar linear potential in addition to the Coulomb-like short-distance part, a form formulated and used long ago, independently of the present problem. We also use the improved lattice computations discussed in the companion paper~\cite{emmanuel}, including a larger spatial extension, better accuracy, and extending the computation to the axial distributions for both the ground states [$j^P=(1/2)^-$] and the nearest orbital excitations [$j^P=(1/2)^+$].~\footnote{Throughout this paper we will use the heavy quark spectroscopic notation $j^P=(1/2)^-$ which stands for the ground state doublet of heavy-light mesons $J^P= [0^-, 1^-]$, while the lowest excited states with $L=1$, correspond to $j^P=(1/2)^+$, i.e. to the doublet of heavy-light mesons $J^P= [0^+, 1^+]$.  }

As we will show, the two types of models chosen in this paper give similar results, and are well compatible with the lattice data. In particular the (vector) charge distribution for $(1/2)^-$ states is now very well reproduced, while the comparison between the lattice results and the distributions deduced from the quark models is more than satisfactory not only for the $(1/2)^-$ doublet, but also for their excited counterparts with $j^P=(1/2)^+$.
We should also emphasize that in this paper we restrict our attention to elastic matrix elements: we do not consider distributions of the transition matrix elements between states belonging to different doublets of heavy-light mesons, or between radially excited and ground states (for the spectrum of such states please see ref.~\cite{blossier0911.1568}~\footnote{The authors of that paper also plan to compute the transition matrix elements of the axial charge between radially excited and ground heavy-light static mesons.}).

\section{Two proposed quark models}

We will confront the lattice data with results obtained by using a Dirac-type quark model, and one based on the Bakamjian--Thomas (BT) approach~\cite{bakamjian}. We advocate them both, because both exhibit a large set of interesting features and successful results. Although their respective structure seems rather different, they share several common features. 

First,  both quark models are relativistic, i.e. they have a relativistic kinetic energy. By now, there is no doubt that in contrast to heavy quarkonia, heavy-light systems are highly relativistic, necessitating at least a partially  relativistic treatment of kinetic energy~\cite{mormor}. In the two quark models we consider here, this is made by (a) using the standard Dirac equation, or by (b) introducing the free kinetic energy of a quark, $\sqrt{p^2+m^2}$, as in the well-known potential model by Godfrey and Isgur (GI)~\cite{godfrey}. Other relativistic features come from the treatment of the potential as having a Lorentz structure of  vector or scalar exchange in the space of Dirac spinors.

Second, they are based on the standard assumption of a linear confining (scalar) potential, plus a Coulomb-like short-distance part.  We are of course aware that many objections have been formulated against a scalar confinement potential. Phenomenologically, however, it offers a simple and satisfactory description of the data, for which there exists no real alternative.~\footnote{Recently, a new phenomenological objection has been formulated by the Wisconsin group~\cite{allen}. They predict a rapid variation of the excitation energy with the light quark mass in a heavy-light system, which would contradict the experimental data. We find this to be a consequence of the fact that they assume a hard Coulomb potential, $\propto 1/r$. In fact their prediction changes completely after softening this $1/r$ behavior, as should be done according to asymptotic freedom. As a result one finds that the excitation energy is slowly varying with the light quark mass, as observed.}

Other than these common features, the two models are rather different, and so are the expressions for hadronic matrix elements, which makes the agreement of the physical results obtained from both  quite surprising (see sec.~\ref{comparaison} below). Before entering the specifics of both models, we should add a few comments.
\begin{itemize}
\item[(i)]\underline{The meaning of the Coulomb-like potential:} As usual, a Coulomb-like potential is introduced to mimic physics at ``short-distances". It is, however, unclear what that assumption exactly means. It should be stressed that the naive idea of one-gluon exchange dominating at short-distance is not a sufficient justification for the assumption.  Indeed, studies of the static potential by using lattice QCD~\cite{bali} suggest that the one-gluon exchange can dominate only at very small distances, hardly accessible from the lattice data, and only a small part of the static potential is describable by perturbative QCD, even after including the perturbative higher order corrections.
On the whole, the Coulomb-like potential in phenomenological models covers a large range of distances and should be probably considered only as a phenomenological description that, together with the linear contribution, provides the medium-range potential responsible for the bound states. This is manifested by the fact that $\kappa=4 \alpha_s/3$ is not necessarily small. In the Dirac model, as well as in the Cornell model of quarkonia, it is in fact large.
\item[(ii)]\underline{Interest of the Dirac-type models:} We may say that the Dirac equation formalism has two main features which fit particularly well the present problem.
First, it corresponds to the static limit, which is the one considered on the lattice side. Second, the expressions for various spatial distributions of densities is particularly simple; they are just quadratic combinations of the upper and lower components of the radial Dirac wave functions. 
A phenomenological  drawback of the static limit is related to the fact that the spectrum of states is not the physical one, and one cannot directly rely on the physical spectrum to fix the model parameters. One instead has to rely on  extrapolations from the sectors with a heavy $c$- or $b$-quark to the static limit. Moreover, the experimental checks of the predictions made in the static heavy quark limit can be at best made in these heavy-light systems.
\item[(iii)]\underline{Interest of the Bakamjian--Thomas approach:}  This approach allows one to formulate relativistic states in motion, for any mass of all the quarks. In the BT-type models one has an exact representation of the Poincar\'e group. However, it is not known in this scheme how to define fully covariant current operators. Interestingly, it was found that, in the heavy quark mass limit, and the current acting on the heavy quark, covariance of current matrix elements is obtained, as also Isgur-Wise scaling  ~\cite{5r}. In the case of finite mass, and in the present case of density distributions, where the active quark is the light quark, full covariance is lost. However, we hope nevertheless that important relativistic effects are still incorporated in the model. The BT scheme considerably enlarges the domain of physical checks that can be made, although here we work in the static limit. In addition, we adopt as the rest frame wave equation the relativistic wave equation derived by GI~\cite{godfrey}, which in our opinion is unique in having been tested over a particularly large set of states and reactions. 

\par \vskip 0.5 truecm

Concerning the parameters of both the Dirac and BT models, we must first say that neither of them has been fitted from the spatial distributions themselves, that constitute predictions of the two schemes. In both models the parameters have been fixed to describe the hadron mass spectrum only. In the case of the BT model we adopt the original parameters of the Godfrey and Isgur (GI) potential~\cite{godfrey}.

\end{itemize}

\subsection{A model with Dirac equation}

Our Dirac model is, of course, based on the equation~\cite{dirac1999} 
\bea
\biggl[\ \vec \alpha \cdot \vec p \ +\ m \beta \ +\ V(\vec r)\ -\ E \biggr]\, \Psi (\vec r) = 0\,.
\label{DEQ}
\eea

The differences between the eigenvalues $E$ correspond to the differences between the hadron masses, and are used to fix the parameters of the potential $V(\vec r)$, which is discussed below. Once these parameters are fixed, we only deal with eigenfunctions $\Psi (\vec r)$. From the knowledge of the quadrispinor $\Psi (\vec r)$, one computes straightforwardly the density distributions, which are defined by
\bea
\rho_{\cal O} (\vec r)= \Psi (\vec r)^{\dagger}{\cal O} \Psi (\vec r) \,,
\eea
with the desired Dirac matrix being ${\cal O}=1, \vec \sigma, \beta$, for vector ($V$), axial ($A$), and scalar ($S$) density, respectively. One must also take into account the angular momentum  $j$  of the light quark and of the heavy quark spin, to form the desired $J^{P}$ - eigenstates. It is then very easy to establish, by purely algebraic calculation, the explicit expressions for the various matrix elements of local densities for the ground states, in terms of radial wave functions  $f_{1/2}^{-1}(r)$ and $g_{1/2}^{-1}(r)$, which enter the full wave function as 
\bea
\hspace*{-3mm}\phantom{\Huge{l}} \raisebox{6mm}{\phantom{\Huge{j}}}
\Psi_{jm}^{k}({\vec r})&=&
			\left(\begin{array}{c}
\hspace*{-3mm}\phantom{\Huge{l}} \raisebox{-.2cm}{\phantom{\Huge{j}}}
f_{j}^{k}(r)\ {\cal Y}_{jm}^{k}({ \widehat{r}})\, \\
\hspace*{-3mm}\phantom{\Huge{l}} \raisebox{.0cm}{\phantom{\Huge{j}}}
ig_{j}^{k}(r)\ {\cal Y}_{jm}^{-k}({ \widehat{r}})\, 
\end{array}\right) \,.§
\eea
\vspace*{.1cm}\\
\noindent
Notice that ${\cal Y}_{jm}^{-k}({ \widehat{r}})$ denotes the combination of spherical harmonics with spinors standardly used in the solution of the Dirac equation, with the definitions of Landau and Lifshitz~\cite{landau}: $k$ is the Dirac quantum number [$k=l$ for $l=j+\frac{1}{2}$, and $k=-(l+1)$ for $l=j-\frac{1}{2}$], which is $k=-1$  ($k=1$)  for the $J^P=0^{-}$ ($0^{+}$) state, and $k=-2$ for $J^P=2^{+}$. After inserting this into the Dirac equation, and integrating over the angular variables, one arrives at a system of coupled differential equations for the radial functions $f_{j}^{k}(r)$, $g_{j}^{k}(r)$, that can be computed numerically by a method of expansion over a basis of orthogonal functions explicitly given in the Appendix  of the present paper. In what follows, the results are obtained in the basis of dimension $n = 40$.~\footnote{In the Appendix we provide the reader willing to reproduce our results with expansion coefficients in the basis of dimension $n=15$. Extension to $n=40$ is obviously straightforward. If necessary the coefficient in the basis with $n=40$ can be obtained upon request from the authors.} The details of such a calculation have been already described in detail in ref.~\cite{dirac1999} where we obtained the expression for the matrix element of the axial current. Here we give the results for all three matrix elements:
\bea\label{dirac-distr-0}
\rho_{V}(r)=\bigl| f_{1/2}^{-1}(r)\bigr|^2+\bigl| g_{1/2}^{-1}(r)\bigr|^2 \,,\nonumber \\
\rho_{A}(r)= \bigl| f_{1/2}^{-1}(r)\bigr|^2-{1\over 3}~\bigl| g_{1/2}^{-1}(r)\bigr|^2\,, \\
\rho_{S}(r)=\bigl|  f_{1/2}^{-1}(r)\bigr|^2-\bigl| g_{1/2}^{-1}(r)\bigr|^2\,,\nonumber
\eea
for the vector, axial, and scalar density distributions, respectively.~\footnote{Note that the vector and scalar densities are often referred to as {\it charge} and {\it matter} densities, respectively.} 
As mentioned above, these expressions are simple quadratic combinations of the radial wave functions. One must warn that, in the case of the axial current operator $\vec \sigma$, since it is a space-vector, one must consider the transition between $1^{-}$ and $0^{-}$ states. One could define various distributions according to the direction of $\vec \sigma$. We choose the one calculated
in the companion paper \cite{emmanuel}, which corresponds to the emission or absorption of the pion between the $0^{-}$ and the $1^{-}$. Moreover, there are sign conventions to be fixed (emission or absorption, relative phase of the two states). The above one is in agreement with what has been chosen in ref.~\cite{emmanuel}.  It is trivial to see that there is a linear relation among the above three distributions, namely,
\bea\label{identity_mat}
\rho_{S}(r)=\frac{1}{2} \left[ 3\rho_{A}(r)-\rho_{V}(r)\right]\,, 
\eea
that is also valid in the Bakamjian--Thomas approach, as we show in the next subsection. This feature is manifest in models but not in QCD where the scalar density is renormalization scheme and scale dependent, in contrast to the vector and axial ones.  With that in mind we will also consider the scalar density distribution and check the extent to which the above relation is valid (up to a proportionality constant). 
\vskip 0.1cm
Similar quantities for the $j^P=(1/2)^+$ doublet have been computed on the lattice too~\cite{emmanuel}. The corresponding quark model expressions are:
\bea\label{dirac-distr-1}
\rho^+_{V}(r)=\bigl| f_{1/2}^{1}(r)\bigr|^2+\bigl| g_{1/2}^{1}(r)\bigr|^2 \,,\nonumber \\
\rho^+_{A}(r)= \bigl| g_{1/2}^{1}(r)\bigr|^2-{1\over 3}~\bigl| f_{1/2}^{1}(r)\bigr|^2\,, \\
\rho^+_{S}(r)=\bigl|  f_{1/2}^{1}(r)\bigr|^2-\bigl| g_{1/2}^{1}(r)\bigr|^2\,,\nonumber
\eea 
where the superscript $+$ denotes parity and is used to distinguish from the expressions given in eq.~(\ref{dirac-distr-0}).~\footnote{To avoid ambiguities, when we discuss the features applicable to both the ground and excited states, we will denote them by a superscript  $(+)$, e.g.  $\rho^{(+)}_{V}(r)$ }  Like in the case of ground states, also these densities in quark modes satisfy a simple relation,
\bea\label{identity_matp}
\rho^+_{S} (r)=\frac{1}{2} \left[  \rho^+_{V}(r)-3~ \rho^+_{A}(r)\right]\,. 
\eea
To obtain the above radial expression in the axial case requires consideration of the transition between $1^+$ and $0^+$ heavy-light mesons, once more because the operator is a space-vector. 
The convention chosen here for the axial distribution is opposite to the one in the lattice paper. We take this into account in our comparison below, by converting the sign of the lattice data. 

\begin{figure}[t!]
\label{fig:1}
\begin{center}
\hspace*{-11mm}\resizebox{85mm}{!}{\includegraphics{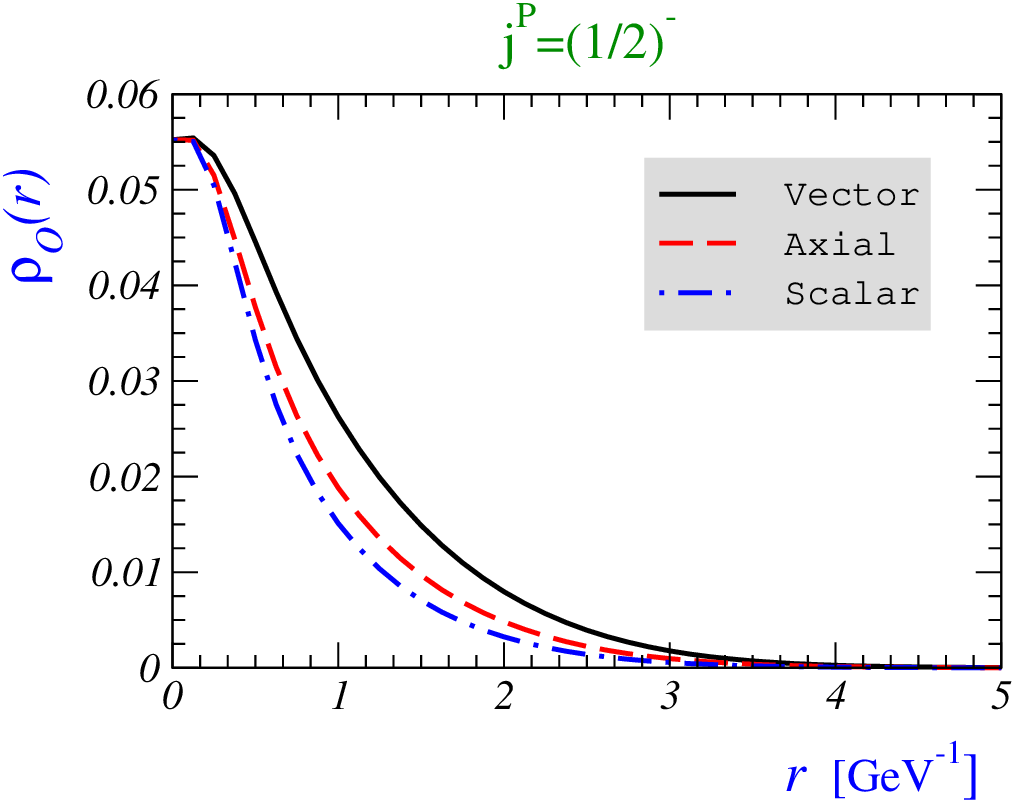}}
%\vspace*{8mm}
\resizebox{85mm}{!}{\includegraphics{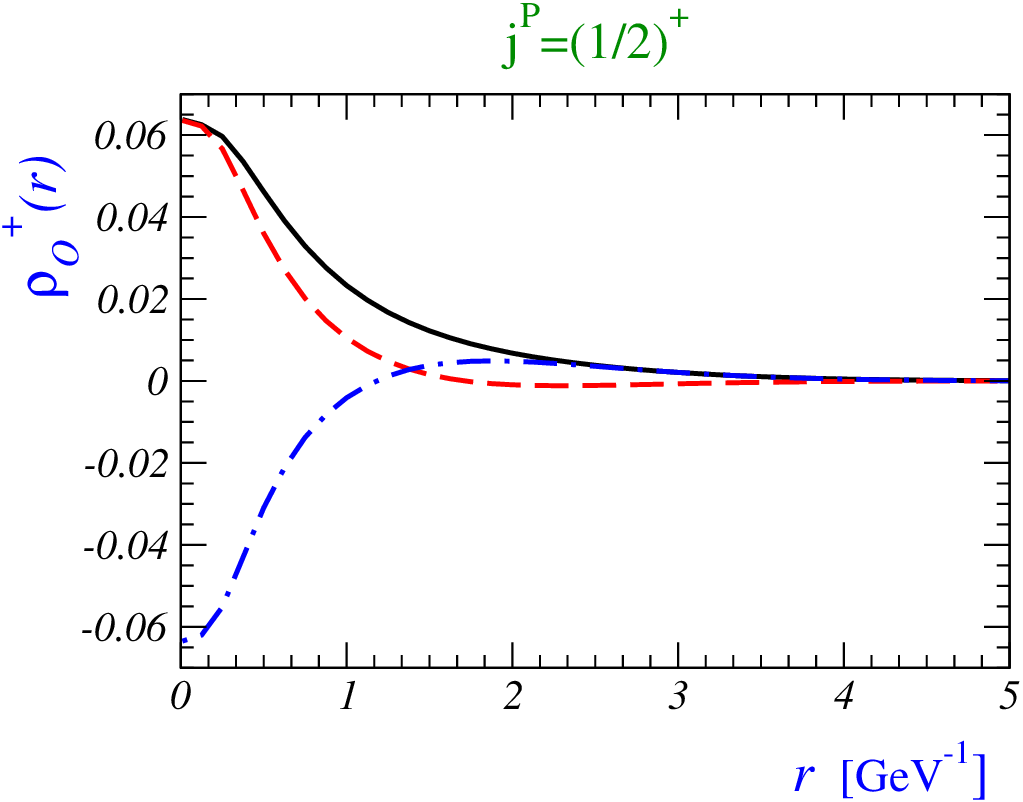}}
\caption{\footnotesize{\sl 
Radial distributions for the light-light vector and axial current, as well as for the scalar density, in the case in which the external states belong to the lowest lying doublet of static heavy-light mesons [$j^P=(1/2)^-$] is presented in the left plot, while in the right one the external states are the static heavy-light mesons belonging to the nearest orbital excitations  [$j^P=(1/2)^+$]. 
The results are obtained by using the Dirac-type model discussed in the text. }} 
\end{center}
\end{figure}

At this point we would like to underline the fact that the vector density is of course positive definite, while the axial and scalar densities are not. In fact they are expressed in terms of differences between the large and small component contributions [see eqs.(\ref{dirac-distr-0}, \ref{dirac-distr-1})], and therefore can change sign as a function of $r$. This actually happens for $j^P=(1/2)^+$ states as we show below, and clearly demonstrates the crucial importance of the relativistic treatment since in the nonrelativistic approximation the small component vanishes. See Fig.~\ref{fig:1} where the three densities obtained from the Dirac model are compared with those obtained from lattice QCD.

One should be particularly careful in choosing the orthogonal basis to diagonalize the Hamiltonian when dealing with $(1/2)^+$ states. It was proposed in ref.~\cite{olsson} to use the pseudo-Coulombic (PC) basis, and useful formulas were provided. These formulas were also used in ref.~\cite{dirac1999}. Near the origin the basis functions behave as $\sim r^{\ell_{PC}}$. However, in \cite{olsson} it was assumed that the orbital angular momentum $\ell_{PC}$ of the basis is identical to the $\ell$ of the Dirac equation ($\ell=j \pm 1/2$) which gives the correct behavior for the upper components, with a regular potential. However, the lower components behave differently, as $r^{\ell'}$, where $\ell'=\ell \mp 1$.
In the case $L=1,j=1/2$, the $\ell$ of the Dirac equation $\ell =L=1$, but $\ell'=0$, and therefore the $\ell_{PC}$ of the basis functions for the lower components should be now chosen to be $\ell_{PC}=0$. This has been implemented in the present paper. The main effect is that the above densities do not tend to zero, having always a contribution from the lower components. Our previous results reported in ref.~\cite{dirac1999} for the spectrum and the matrix elements remain essentially unchanged.

\subsubsection{Parameters of the Dirac model} \label{parameters} The remaining problem is then to determine the potential and the mass in eq.~(\ref{DEQ}). We choose the following form of the potential:
\bea\label{pot-dir}
V(r)=\beta \left( a r+c \right)  -\kappa {d(r)\over r}\,,\qquad {\rm with}\quad d(r)=1-e^{-(r/R)^2}\,.
\eea
The function $d(r)$ in the vector part of the potential is chosen to temper the Coulombic singularity at the origin, and its introduction does not alter the discussion made in our previous paper in which we took $d(r)=1$~\cite{dirac1999}.  As for the scalar part of the potential its form is the same as in ref.~\cite{dirac1999}, where we discussed the puzzling discrepancy between  the QCD sum rule prediction of $\widehat{g}$ and its much larger experimental value, 
\bea
\widehat{g}\equiv g_A=\int d^3\vec{r}~\rho_{A}(r)\,.
\eea  
Notice in particular that the constant $c$ in eq.~(\ref{pot-dir}) combines with the constituent quark mass into a constant term which does not have to be positive.  
Since our paper~\cite{dirac1999}, new information concerning the spectrum of heavy-light mesons has been obtained both in experiment and through numerical simulations of QCD on the lattice. This is why we will somewhat readjust our set of {\it preferred parameters}  given in ref.~\cite{dirac1999}. The most important constraint on the model parameters comes from $\Delta E_1 = m_{(L=1)}-m_{(L=0)} \approx 0.4$~GeV. 
Because we are working in the heavy quark limit, we should in principle distinguish the $(1/2)^+$ states  from  the $(3/2)^+$ ones, both being $L=1$. The mass degeneracy of these states in the heavy quark limit is the second important constraint on the choice of model parameters. We therefore require the spin-orbit splitting to be negligible, in agreement with the celebrated model by Godfrey and Isgur~\cite{godfrey}. 
In the Dirac model this implies, with the known value of the string tension, $a$, that the Coulombic term is rather large, far from any perturbative regime. 
This is not a problem, in our opinion, since we do not consider it as a short-distance one-gluon term, but rather as an effective contribution at medium distances.

As already mentioned, only the combination 
\bea\label{eq8}
c_0=m+c\,,
\eea 
enters the Dirac equation and it may be negative. From experience with potential models, we know that the confining potential is most often written as  $a r+c$, with $c$ being negative and large. Therefore, one should not object to $c_0 <0$, as it results from our fits. In ref.~\cite{dirac1999} we found $c_0=-0.2$~GeV, which was an important distinctive feature of our model. As we explain below, in this paper we find $c_0=-0.1$~GeV, still negative. The fact that $c_0<0$ also implies an important cancellation of the linear term of the potential in the region around $r \simeq 1~\gev^{-1}$, for $a \simeq 0.2~\gev^2$, which will play an important role in the apparent approximate $\gamma_5$-symmetry discussed in sec.~\ref{comparaison}. 

In the following we discuss in more detail the data used to determine the parameters in the potential~(\ref{pot-dir}).
\begin{itemize}
\item {\bf Experiment}:  The experimental observations of $D$ and $B$ states with $L=1$ [$j=(3/2)^+$]~\cite{B**,PDG} allow us to linearly extrapolate to the static heavy quark mass limit where we find $\Delta E_1\simeq 0.4$~GeV, for both strange and nonstrange heavy-light mesons. 
\item  {\bf Lattice QCD}:  A recent lattice QCD study of the spectrum of heavy-light mesons in the static limit reports $\Delta E_1\simeq 0.4~$GeV, and indicates the spin-orbit splitting to be negligible, both in quenched approximation and after including $N_F=2$ dynamical light quarks~\cite{burch0809.1103}.~\footnote{Note that the conclusions of the ETM Collaboration are different. Their first paper~\cite{wagnerETMC1}  is superseded by the second one~\cite{wagnerETMC2}, which interprets differently the strange state, by a ``partially" quenched calculation where the sea is constituted by $u/d$ loops. In ref.~\cite{wagnerETMC2}, the spin-orbit mass difference is the same for strange and nonstrange states, i.e. $0.1$~GeV, thus larger than the negligible one found in ref.~\cite{burch0809.1103}. Note that the calculations of ref.~\cite{burch0809.1103} are ``{\it unitary}", i.e. the sea and the valence light quarks are degenerate. In the case of the non strange mesons, the ETMC data are also unitary, so the two results are comparable but do not agree. We have no explanation for this discrepancy.} 
\end{itemize}
Notice, however, that our assumption of small spin-orbit splitting has been challenged in the $D_s$ sector: $m_{D_{s2}^\ast} - m_{D_{s0}^\ast}\simeq 0.25$~GeV.~\footnote{Note that such a large value is also found  in the case of non-strange $D$ mesons: $m_{D_2^\ast} - m_{D_{0}^\ast}\simeq 0.14(3)$~GeV.}  Presumably, what is observed experimentally for the $D_{sJ}$ states, i.e. very large spin-orbit splittings, is a finite mass effect, although that phenomenon is not easy to understand in quark models. 

Another important spectroscopic information is a newly identified radial excitation of $D_s$~\cite{D_s'}.  The radial excitation energy has been determined cleanly on the lattice too. In the static limit, in quenched approximation and for the light quark around the strange quark mass, it is found to be around $\Delta E' \simeq 0.6$~GeV~\cite{blossier0911.1568},  in qualitative agreement with experiment and with quark models for the $D_s$.  Similar methods applied to  $N_F=0$ and $N_F=2$ lattice data~\cite{burch0809.1103} give a similar value.~\footnote{It is important to mention that the conversion from lattice to physical units in both ref.~\cite{blossier0911.1568} and ref.~\cite{burch0809.1103} has been made by using $r_0=0.5~$fm.}  In summary, the radial excitation is found to be higher than the $L=1$ states.

We emphasize that the parameters used in this paper are not far from the set of parameters used in ref.~\cite{dirac1999}. We have in fact tried to remain as close as possible to this old set of parameters, while maintaining the assumption of a very small spin-orbit splitting, which leads us to the following modifications:

(i) To increase $\Delta E_1$ while maintaining the  spin-orbit splitting small requires a simultaneous increase of the string tension $a$, and of the strength of the Coulombic term $\kappa$, with respect to the values we obtained in ref.~\cite{dirac1999}. We also choose to respect the mass difference between the strange and nonstrange heavy-light mesons, $\simeq 0.100~\gev$, to obtain: $a=0.25~\gev^2$, $\kappa=0.75$, with $d(r) \neq 1$ only at very small distance. The constant $c_0$ is found to be about $c_0=-0.1~\gev$ for the strange, and $c_0=-0.35~\gev$ for the nonstrange states.

(ii) The choice of function $d(r)$, introduced in eq.~(\ref{pot-dir}), is meant to maintain a small  splitting between $(3/2)^+$ and $(1/2)^+$  states, independently of the light quark mass.  At small distances $d(r) \ll 1$  allows one to kill the Coulomb singularity. With a strict Coulomb-like potential, and large $\kappa$, we would have obtained  the spin-orbit splitting increasing too rapidly with the light quark mass, which contradicts the observations made in lattice QCD~\cite{burch0809.1103}. 
Indeed the larger the light quark mass, the more the wave function is concentrated at the center and the weight of the Coulomb piece in the potential increases. As a consequence the contribution of the vectorlike part to spin-orbit splitting exceeds the scalarlike one, and the spin-orbit splitting grows with the light quark mass. To suppress that effect, it suffices to cut off the singularity, which we do by introducing a Gaussian cutoff, namely a factor $d(r)=1-\exp[-(r/R)^2]$. We choose $R \simeq 0.25~\gev^{-1}$, a rather short-distance. This cut is enough to keep the spin-orbit splitting very small independently of the light quark mass. 
A similar procedure to soften the Coulomb singularity has been used in the GI model.

With the above set of parameters and for the strange light quark we obtain   $\Delta E_1 \simeq 0.4$~GeV  and for the spin-orbit splitting $E_{P_{3/2}}-E_{P_{1/2}}=-0.01~$GeV, as expected. 
We then deduce the mass difference between the radial excitation and the ground state to be $\Delta E^\prime = 0.576$~GeV, in agreement with lattice and experimental findings. 
For nonstrange states we have $\Delta E_1 \simeq 0.38$~GeV  and  $E_{P_{3/2}}-E_{P_{1/2}}=0.01~$GeV, and predict $\Delta E^\prime = 0.556$~GeV. The difference between the strange and nonstrange 
ground states is $0.1~$GeV.

\subsection{The Bakamjian--Thomas approach to relativistic quark models}
 
In this Section we compute the same distributions of heavy-light mesons but in the class of  Bakamjian--Thomas (BT) relativistic quark models~\cite{bakamjian,2r,3r,4r,5r}. 
This is a class of models with a fixed number of constituents where the states are covariant under the Poincar\'e group. 
The model relies on an appropriate Lorentz boost of eigenfunctions of a Hamiltonian describing the spectrum at rest. As it has been explicitly shown, the model has the important properties that the matrix elements of currents between hadrons are covariant in the heavy quark limit and exhibit Isgur--Wise scaling~\cite{5r,6r,7r,8r}. Given a Hamiltonian describing the spectrum, the model provides an unambiguous result for the elastic Isgur-Wise function $\xi (w)$ \cite{5r,9r}, and satisfies several sum rules (SR) in the heavy quark limit of QCD, such as Bjorken or Uraltsev SR, as well as SR involving higher derivatives of $\xi (w)$ at zero recoil~\cite{10r,11r}.

One can compute the $V$, $S$, and $A$ distributions for mesons within the BT scheme using a Hamiltonian describing the spectrum, like the GI model \cite{godfrey}, that has been used elsewhere~\cite{6r} to compute the elastic Isgur-Wise function $\xi (w)$ and the inelastic ones $\tau_{1/2}(w)$, $\tau_{3/2}(w)$ \cite{9r}. Since the computation of the light-light matrix elements within the BT scheme has not been presented before, we present it here in more detail.\par

\hspace*{\parindent}
\subsubsection{The Bakamjian--Thomas approach to quark models}
\hspace*{\parindent}
We first recall some basic elements of the BT-approach to quark models. The construction of the BT wave function in motion involves 
a {\it unitary} transformation that relates the wave function $\Psi_{s_1, \cdots ,s_n}^{(P)}({\vec{p}}_1, \cdots , {\vec{p}}_n)$ in terms of one-particle variables, the spin $\vec{ S}_i$ and momenta $\vec{p}_i$ to 
the so-called {\it internal} wave function $\Psi_{s_1, \cdots ,s_n}^{int}(\vec{P}, \vec{k}_2, \cdots ,\vec{k}_n)$ given in terms of another set of variables, the total momentum $\vec{P}$ and the internal momenta $ \vec{k}_1, \vec{k}_2, \dots ,\vec{k}_n$ ($\sum\limits_i \vec{k}_i = 0$)~\cite{5r}. This property ensures that, starting from an orthonormal set of internal wave functions, one gets an orthonormal set of wave functions in any frame. The basis $\Psi_{s_1, \cdots , s_n}^{(P)}(\vec{p}_1, \cdots , \vec{p}_n)$ is useful to compute one-particle matrix elements like current one-quark matrix elements, while the second  $\Psi_{s_1, \cdots ,s_n}^{int}(\vec{P}, \vec{k}_2, \cdots ,\vec{k}_n)$ allows one to exhibit Poincar\'e covariance. In order to satisfy the Poincar\'e commutators, the unique requirement is that the mass operator $M$, i.e. the Hamiltonian describing the spectrum at rest, should depend only on the internal variables and be rotationally invariant, i.e. $M$ must commute with $\vec{P}$, ${\partial \over \partial \vec{P}}$ and $\vec{S}$. The internal wave function at rest $(2 \pi )^3 \delta (\vec{P}) \varphi_{s_1, \cdots , s_n}(\vec{k}_2, \cdots , \vec{k}_n)$ is an eigenstate of $M$, $\vec{P}$ (with $\vec{P} = 0$), $\vec{S}^2$ and $\vec{S}_z$,  while the wave function in motion of momentum $\vec{P}$ is obtained by applying the boost $\vec{B}_P$, with $P^0 = \sqrt{\vec{P}^2 + M^2}$ involving the dynamical operator $M$.

The final output of the formalism that gives the total wave function in motion, $\Psi_{s_1, \cdots ,s_n}^{(P)}$, in terms of the internal wave function at rest $\varphi_{s_1, \cdots , s_n}(\vec{k}_2, \cdots , \vec{k}_n)$ is encoded in the following formula
\bea
\label{1e}
\Psi_{s_1, \cdots ,s_n}^{(P)}(\vec{p}_1, \cdots , \vec{p}_n) &=& (2\pi )^3 \delta \left ( \sum_i \vec{p}_i - \vec{P}\right )  \sqrt{{\sum\limits_i p_i^0 \over M_0}}  \left ( \prod_i {\sqrt{k_i^0} \over \sqrt{p_i^0}}\right )\nn \\
&&\times \sum_{s'_1, \cdots , s'_n} \left [ D_i({\bf R}_i)\right ]_{s_i,s'_i} \varphi _{s'_1, \cdots , s'_n} (\vec{k}_2, \cdots , \vec{k}_n)\,,
\eea
where $p_i^0 = \sqrt{\vec{p}_i^2 + m_i^2}$, and $M_0$ is the free mass operator, given by $M_0 = \sqrt{(\sum\limits_i p_i)^2}$.

The internal momenta of the hadron at rest are given in terms of the momenta of the hadron in motion by the {\it free boost} $k_i = {\bf B}_{\sum\limits_i p_i}^{-1} p_i$ where the operator ${\bf B}_p$ is the boost $(\sqrt{p^2}, \vec{0}) \to p$, the Wigner rotations ${\bf R}_i$ in the preceding expression ${\bf R}_i = {\bf B}_{p_i}^{-1}{\bf B}_{\sum\limits_i p_i}^{-1} {\bf B}_{k_i}$ and the states are normalized according to $\langle \vec{P'}, {S'}_z\vert \vec{P}, {S}_z\rangle \ = \ (2 \pi )^3 \delta ( \vec{P'} - \vec{P}) \delta_{S_z,S'_z}$.\par

The current one-quark matrix element acting on quark 1 between two hadrons is then given by the expression
\bea
\label{2e}
\langle \Psi ' (\vec{P'}, {S}'_z)| J^{(1)}| \Psi (\vec{P}, {S}_z)\rangle &=& \int {d \vec{p'}_1 \over (2 \pi)^3}\ {d \vec{p}_1 \over (2 \pi)^3} \left ( \prod_{i=2}^n {d \vec{p}_i \over (2 \pi)^3}\right )\nn \\
&&\hspace*{-35mm}\Psi_{s'_1, \cdots ,s_n}^{P'} (\vec{p'}_1 , \cdots , \vec{p}_n)^* < \vec{p'}_1, {s'}_1|J^{(1)}|\vec{p}_1, {s}_1>\Psi_{s_1, \cdots , s_n}^{P} (\vec{p}_1 , \cdots ,\vec{p}_n),
\eea
where $\Psi_{s_1, \cdots , s_n}^{P} (\vec{p}_1 , \cdots , \vec{p}_n)$ is expressed via the internal wave function as in eq.~(\ref{1e}), and $\langle \vec{p'}_1, s'_1\vert J^{(1)}\vert \vec{p}_1, s_1\rangle$ is the one-quark current matrix element.\par

As demonstrated in refs.~\cite{5r,7r}, in this formalism,  current matrix elements in the heavy quark limit are covariant, they exhibit Isgur-Wise scaling, and one can compute Isgur-Wise functions like $\xi (w)$, $\tau_{1/2}(w)$, $\tau_{3/2}(w)$~\cite{6r}.

In the present paper, we deal with light quark currents between heavy-light meson states, and there is no {\em a priori} reason for the result to be covariant. To compute the desired radial distributions we need to specify the mass operator $M$, which we choose to be the one appearing in the GI model, and we work in the meson rest frame.

\hspace*{\parindent}
\subsubsection{Ground state vector, scalar, and axial densities}
\hspace*{\parindent}
We need here to compute the matrix elements of the operators $O=\overline{q}\Gamma{q}$ where $\Gamma$ is any Dirac matrix, in practice $\Gamma=\gamma^0, 1, \vec{\gamma}\gamma^5$, for the vector current, scalar, and axial vector current densities respectively.  In ref.~\cite{5r} we did compute matrix operators of the type (\ref{2e}) where we took the convention ``1" for the heavy and ``2" for the light quark. The heavy mass limit was taken for the quark  1. In the case we are treating here, the light quark is coupled to the current, while the heavy mass limit is to be taken for the spectator. We keep  the same notation as in ref.~\cite{5r}, and for finite mass for both quarks the matrix element (\ref{2e}) one writes
\bea
\label{3e}
&&\langle\overline{B}(\vec{P}')|\overline{q}(0)\Gamma q(0)|\overline{B}(\vec{P})\rangle\nn \\
&&= \int{ d\vec{p}_1 \over ( 2 \pi )^3 }{ 1\over p_1^0}\ \varphi (\vec{k}_1)^*\ \varphi (\vec{k'}_1) {\sqrt{u^0u^{'0} \over p^0_2 p^{'0}_2 } } \sqrt{k{_1^0}k{_2^0} \over (k{_1^0}+m_1)(k{_2^0}+m_2)}\sqrt{k'{_1^0}k'{_2^0} \over (k'{_1^0}+m_1)(k'{_2^0}+m_2)} \nn \\
&&{1 \over 16}{\rm Tr} \left [  \Gamma \left( m_2 + {/ \hskip-2.5 truemm p}_2 \right) (1 + {/ \hskip-2.5 truemm u}) \left ( m_1 + {/ \hskip-2.5 truemm p}_1 \right ) (1 + {/ \hskip-2.5 truemm u'}) \left ( m_2 + {/ \hskip-2.5 truemm p'}_2 \right )\right ]\,,
\eea
where  the internal wave function at rest,  $\varphi (\vec{k}_2)$, is normalized as
\bea
\label{4e}
\int {d\vec{k}_2 \over (2 \pi )^3}\ | \varphi (\vec{k}_2)|^2 = 1\,,
\eea
\noi $p_1$, $p'_1$ and $p_2$, $p'_2$ ($m_1$ and $m_2$) are the quark four-momenta (masses) of the heavy and light quarks respectively. The four-vector $u$, $M_0$, and the momenta $k_i$ and $p_i$ are related as
\bea \label{5e}
u = {p_1 + p_2 \over M_0}\,, \qquad\qquad M_0 = \sqrt{(p_1 + p_2)^2}\,,\qquad\qquad  {\bf B}_uk_i = p_i \,\, (i = 1, 2)\,,
\eea
\noi where $k_1$ and $k_2$ are the four-momenta of the quarks in the rest frame of the $B$ meson and ${\bf B}_u$ is the boost associated with the four-vector $u$.~\footnote{As noted above, the BT states are normalized acccording to
$\langle \overline{B}_d(P')| \overline{B}_d(P)\rangle_{\rm BT} \ = (2\pi)^3 \delta (\vec{P} - \vec{P'})$.}
Taking the heavy quark limit ($m_1 \to \infty$) in eq.~(\ref{3e}) amounts to the following replacements:  $M_0 \to m_1 \to m_B$, $u^{(\prime)} \to v^{(\prime)}$ ,  ${p_1^{(\prime)}/m_1} \to v^{(\prime)}$, $ {k_1^{(\prime)0} / m_1} \to 1$, $k_2^{(\prime)0} \to p_2^{(\prime)}\cdot v^{(\prime)}$, where $v$ and $v^\prime$ are the initial and final $B$-meson four-velocities. In the frame in which $\vec{P} = \vec{0},  \vec{P}' = \vec{q}$ one then obtains:\\
\noindent (i) Vector current, 
\bea
\label{6e}
\langle\overline{B}(\vec{q})|\overline{q}(0)\gamma^0 q(0)|\overline{B}(\vec{0})\rangle &=& \int {d\vec{k} \over (2 \pi )^3}\ \varphi (\vec{k}+ \vec{q})^*\ \varphi (\vec{k})\nn \\
&& \nn\\
&&\hspace*{-47mm}\times {(\sqrt {\vec{k}^2+m^2}+m)(\sqrt {(\vec{k}+\vec{q})^2+m^2}+m)+\vec{k}\cdot(\vec{k}+\vec{q}) \over  2\left[  \sqrt {\vec{k}^2+m^2}\sqrt {(\vec{k}+\vec{q})^2+m^2}(\sqrt {\vec{k}^2+m^2}+m)(\sqrt {(\vec{k}+\vec{q})^2+m^2}+m)\right ]^{1\over2}},
\eea
where we write  $m_2 = m$, since the light quark mass is the only mass in the game. The above matrix elements satisfies $\langle\overline{B}(\vec{0})|\overline{q}(0)\gamma^0 q(0)|\overline{B}(\vec{0})\rangle\ = 1$, as expected from gauge invariance.
\noindent (ii) Scalar density,
\bea
\label{7e}
 \langle\overline{B}(\vec{q})|\overline{q}(0)q(0)|\overline{B}(\vec{0})\rangle &=& \int {d\vec{k} \over (2 \pi )^3}\ \varphi (\vec{k}+ \vec{q})^*\ \varphi (\vec{k})\nn \\
 &&\nn \\
&&\hspace*{-47mm}\times {(\sqrt {\vec{k}^2+m^2}+m)(\sqrt {(\vec{k}+\vec{q})^2+m^2}+m)-\vec{k}\cdot(\vec{k}+\vec{q}) \over  2\left[  \sqrt {\vec{k}^2+m^2}\sqrt {(\vec{k}+\vec{q})^2+m^2}(\sqrt {\vec{k}^2+m^2}+m)(\sqrt {(\vec{k}+\vec{q})^2+m^2}+m)\right ]^{1\over2}}\,,
\eea
which leads to the scalar coupling,
\bea
\label{9e}
g_S =\langle\overline{B}(\vec{0})|\overline{q}(0)q(0)|\overline{B}(\vec{0})\rangle\,.
\eea

The Fourier transforms of (\ref{6e}) and (\ref{7e}) finally give the desired densities in configuration space  
\bea
\label{10e}
\rho_{V}(\vec{r}) =  \int {{d\vec{q} \over (2 \pi )^3} e^{i\vec{q}.\vec{r}} \langle\overline{B}(\vec{q})|\overline{q}(0)\gamma^0q(0)|\overline{B}(\vec{0})\rangle}\,,\quad 
\rho_{S}(\vec{r}) =  \int {{d\vec{q} \over (2 \pi )^3} e^{i\vec{q}.\vec{r}} \langle\overline{B}(\vec{q})|\overline{q}(0)q(0)|\overline{B}(\vec{0})\rangle} \,.
\eea
The densities depend only on the modulus $r = |\vec{r}|$ and, after a lengthy integration over the angles,  one gets
\bea
\label{12e1}
&&\rho_{V}(r) = \rho_1(r) + \rho_2(r)\,,\quad\quad \rho_{S}(r) = \rho_1(r) - \rho_2(r)\,,
\eea
where
\bea
\label{12e}
\rho_1(r) = \left({1\over{2\sqrt{2}\pi^2}} {1\over{r}}  \int_0^{\infty} dq \ q\ \varphi (q) {\sin(qr)(\sqrt {q^2+m^2}+m) \over \left[ \sqrt {q^2+m^2}(\sqrt {q^2+m^2}+m)\right ]^{1\over2}}\right)^2\,,
\eea
\bea
\label{13e}
\rho_2(r) = \left({1\over{2\sqrt{2}\pi^2}} {1\over{r^2}}  \int_0^{\infty} dq \ q\ \varphi (q) {[\sin(qr)-qr\cos(qr)] \over \left[ \sqrt {q^2+m^2}(\sqrt {q^2+m^2}+m)\right ]^{1\over2}}\right)^2\,,
\eea
both square-functions of $r$, which is expected because (\ref{6e}) and (\ref{7e}) are written as a linear combination of two terms, convolution products of the same functions in momentum space.\par

\noindent (iii) Proceeding in the same way, for the matrix element of the axial current
\bea
\label{13ebis}
\langle\overline{B}(\vec{q})|\overline{q}(0)(\gamma\cdot\vec{n}) \gamma^{5} q(0)|\overline{B}^*(\vec{0},\vec{\epsilon})\rangle\,,
\eea
one finds two rotationally invariant terms $\rho_{A}(\vec{r}) = \rho_{A}^{(1)}(r) (\vec{n}\cdot\vec{\epsilon}) + \rho_{A}^{(2)}(r)(\vec{n}\cdot \hat{r}) (\vec{\epsilon}\cdot\hat{r})$,  for which we obtain simple results, related to $ \rho_{S}(r)$ and $ \rho_{V}(r)$ from eq.~(\ref{12e1}),
\bea
\label{13be}
\rho_{A}^{(1)}(r) = \rho_{S}(r)\,,\quad\quad \rho_{A}^{(2)}(r) = \rho_{V}(r)-\rho_{S}(r)\,.
\eea
In configuration space one recovers the normalizations,
\bea
\label{14e}
\int_0^{\infty} dr \ 4\pi r^2\rho_{V}(r) = 1 \,,\quad \quad\int_0^{\infty} dr \ 4\pi r^2\rho_{S}(r) = g_S \,.
\eea
Taking for the axial densities the components such that $\vec{n}=\vec{\epsilon}$, one obtains 
\bea
\rho_A(r) = \rho_{A}^{(1)}(r) + \frac{1}{3} \rho_{A}^{(2)}(r)\,,
\eea
and therefore one verifies the same relation as in the Dirac model, namely
\bea
\label{15ae}
\rho_{A}(r) = {1 \over 3}\ \rho_{V}(r) + {2 \over 3}\ \rho_{S}(r)\,,
\eea
and, by integration over $\vec{r}$,
\bea
\label{15be}
\widehat g = {1 \over 3} + {2 \over 3}\ g_S \,,
\eea
where we use the notation $g_A\equiv \widehat g$.  Notice that relation~(\ref{15ae}) shows that also in the BT scheme there are only two independent functions like in the Dirac model~(\ref{dirac-distr-0}).
\subsubsection{The choice of the mass operator $M$: The Godfrey-Isgur model}
Let us now particularize the above expressions for the choice of the mass operator $M$ given by the GI model \cite{godfrey} that describes the whole set of meson spectra $q\overline{q}$ and $Q\overline{q}$, where $q$ is a light quark ($q$ = $u$, $d$, $s$) and $Q$ is a heavy quark ($Q$ = $c$, $b$), with the important exception of the recently discovered narrow states $D_{sJ}$ $0^+$ and $1^+$, that are too low compared with the predictions of the model. The model contains a relativistic kinetic term of the form
\bea
\label{15ce}
K = \sqrt{\vec{k}^2_1+m{^2_1}} + \sqrt{\vec{k}^2_2+m{^2_2}}\,,
\eea
that is identical to the operator $M_0$ at rest \cite{5r}, and a complicated interaction term that includes : (1) a Coulomb part with a $q^2$ dependent $\alpha_s$, (2) a linear confining piece, and (3) terms describing the spin-orbit and spin-spin interactions. All singularities are regularized - e.g. terms of the type $\delta(\vec{r})$ or $1/m_2$, where $m_2$ is the light quark mass. The Hamiltonian $H$ depends on a number of parameters that are fitted to describe all the meson spectra.
By inserting the GI wave functions (see the Appendix of the present paper) in the above expressions we computed the densities $\rho_{V,A,S}(r)$, and after integrating over $\vec r$, we obtain
\bea
\label{16ae}
g_S \cong 0.56 \,, \qquad \qquad \widehat g \cong 0.71\,.
\eea
Notice that at large distances $r$, $\rho_A(r)$ becomes larger than $\rho_S(r)$, and since this portion of the integral is enhanced by the integration measure growing like $r^2$, the resulting values are different.
For the mean square radii, defined by
\bea
\label{16be}
\langle r^2\rangle_i\ =4 \pi  \int_0^{\infty} dr\ r^4\rho_{i}(r) \qquad (i = V, S, A) \,,
\eea
we obtain 
\bea
\label{16cebis}
\langle r^2\rangle_V\ = 5.174\ \gev^{-2}, \ \ \langle r^2\rangle_S\ = 1.721\ \gev^{-2},  \ \  \langle r^2\rangle_A\ = 2.872\ \gev^{-2},
\eea
that satisfy relation (\ref{15ae}).

\hspace*{\parindent}
\subsubsection{Vector, scalar, and axial current densities for the lowest excited states $(1/2)^+$ (L = 1)}
\hspace*{\parindent}
Like in the discussion with the Dirac-type quark model, we now extend our discussion to the excited heavy-light static mesons with $L=1$, and consider the corresponding transitions $({1/2})^+ \to ({1/2})^+$, i.e. $0^+ \to 0^+$ and $1^+ \to 0^+$.

After proceeding like in the previous subsection, we obtain the following expressions :
\bea
\label{16ebis}
\langle\overline{B}(0^+)(\vec{q})|\overline{q}(0)\gamma^0 q(0)|\overline{B}(0^+)(\vec{0})\rangle \ &=& {1 \over 2} \int {d\vec{k} \over (2 \pi )^3}\ {\varphi^*_{1/2} (\vec{k}+ \vec{q})\ \varphi_{1/2} (\vec{k}) \over |\vec{k}|\ |\vec{k}+ \vec{q}|}\nn \\
&&\times \left[{(\sqrt{\vec{k}^2+m^2}+m)(\sqrt{(\vec{k}+\vec{q})^2+m^2}+m) \over \sqrt{\vec{k}^2+m^2}\sqrt{(\vec{k}+\vec{q})^2+m^2}}\right]^{1/2}\nn \\
&&\hspace*{-21mm}\times  \left[(\sqrt{\vec{k}^2+m^2}-m)(\sqrt{(\vec{k}+\vec{q})^2+m^2}-m)+\vec{k}.(\vec{k}+\vec{q})\right], 
\eea
\bea
\label{17e}
\langle\overline{B}(0^+)(\vec{q})|\overline{q}(0)q(0)|\overline{B}(0^+)(\vec{0})\rangle\
&=& - {1 \over 2} \int {d\vec{k} \over (2 \pi )^3}\ {\varphi^*_{1/2} (\vec{k}+ \vec{q})\ \varphi_{1/2} (\vec{k}) \over |\vec{k}|\ |\vec{k}+ \vec{q}|}\nn \\
&&\times \left[{(\sqrt{\vec{k}^2+m^2}+m)(\sqrt{(\vec{k}+\vec{q})^2+m^2}+m) \over \sqrt{\vec{k}^2+m^2}\sqrt{(\vec{k}+\vec{q})^2+m^2}}\right]^{1/2}\nn \\
&&\hspace*{-21mm}\times \left[(\sqrt{\vec{k}^2+m^2}-m)(\sqrt{(\vec{k}+\vec{q})^2+m^2}-m)-\vec{k}.(\vec{k}+\vec{q})\right],
\eea
which, when Fourier transformed, lead to
\bea
\label{19e}
\rho^+_{V}(r) =  \rho_1^+(r) + \rho^+_2(r)\,,\quad\quad   \rho^+_{S}(r) = -  \rho^+_1(r) +  \rho^+_2(r)\,,
\eea
with $ \rho_1^+(r)$ and $\rho_2^+(r)$ given by the squares,
\bea
\label{19ebis}
\rho^+_1(r)& =& \left[{1 \over {2\sqrt{2}\pi^2}} {1 \over {r}} \int_0^{\infty} dq \ \varphi_{1/2} (q) \sin(qr)(\sqrt {q^2+m^2}-m) \left( {\sqrt{q^2+m^2}+m} \over {\sqrt {q^2+m^2}} \right )^{1\over2} \right]^2,\nn\\
\rho^+_2(r) &= &\left[{1 \over {2\sqrt{2}\pi^2}} {1 \over {r^2}} \int_0^{\infty} dq \ \varphi_{1/2} (q) [\sin(qr)-qr\cos(qr)]  \left( {\sqrt{q^2+m^2}+m} \over {\sqrt {q^2+m^2}} \right )^{1\over2} \right]^2.
\eea
As far as the axial current is concerned, by Fourier transforming the matrix element describing the $1^+ \to 0^+$ transition, 
$\langle \overline{B}(0^+)(\vec{q})|\overline{q}(0)(\gamma\cdot \vec{n}) \gamma^{5} q(0)|\overline{B}^\ast(1^+)(\vec{0},\vec{\epsilon})\rangle $, we find 
\bea
\label{22e}
\rho_{A}^+(\vec{r}) =  \rho_{A}^{+\ (1)}(r) (\vec{n}\cdot \vec{\epsilon}) + \rho_{A}^{+\ (2)}(r)(\vec{n}.\hat{r}) (\vec{\epsilon}\cdot \hat{r}) \,,
\eea
but with relations to $\rho^+_{S}(r)$ and $ \rho^+_{V}(r)$, that are different from eq.~(\ref{13be}), namely,
\bea
\label{23be}
\rho_{A}^{+\ (1)}(r) = -\rho^+_{S}(r)\,,\quad  \quad \rho_{A}^{+\ (2)}(r) = \rho^+_{V}(r)+ \rho^+_{S}(r)\,,
\eea
that finally yield the result for the couplings
\bea
\label{25ae}
\widetilde g = {1 \over 3} - {2 \over 3}\ g_S^+ \,,
\eea
in the notation frequently encountered in the literature, $\widetilde g=g_A^+$. As in the previous subsection, for the numerical calculations we use the GI potential model and, after integrating over $\vec r$, we obtain 
\begin{figure}[t!]
\label{fig:2}
\begin{center}
\hspace*{-11mm}\resizebox{85mm}{!}{\includegraphics{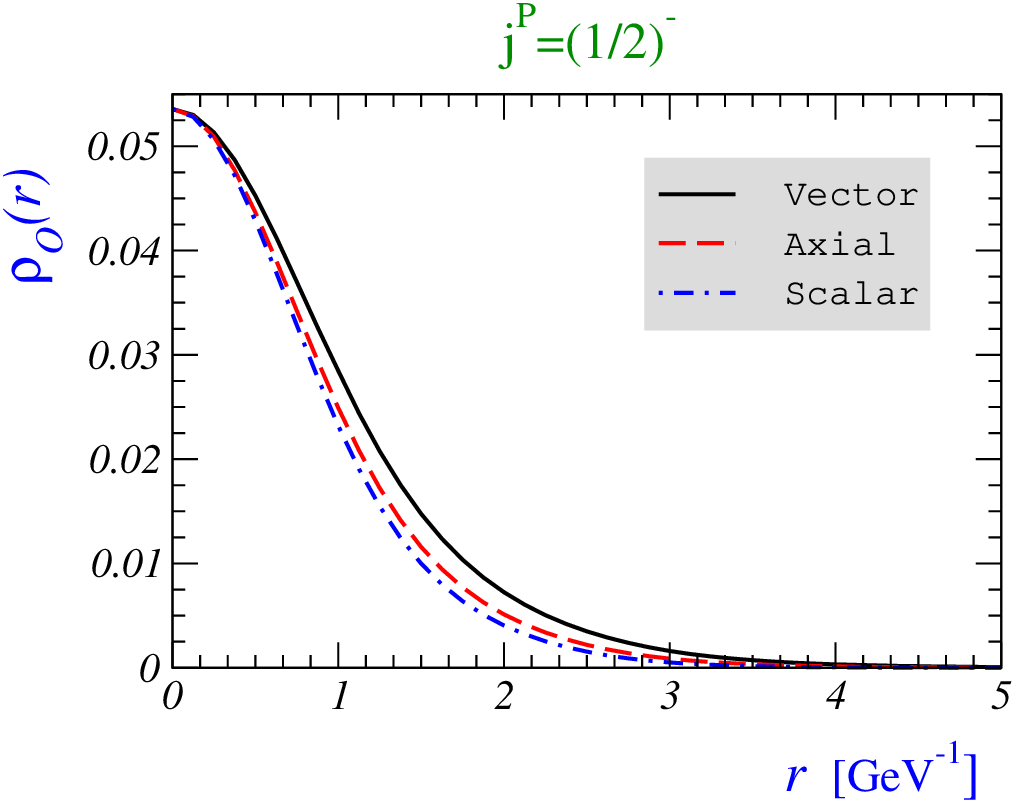}}
%\vspace*{8mm}
\resizebox{85mm}{!}{\includegraphics{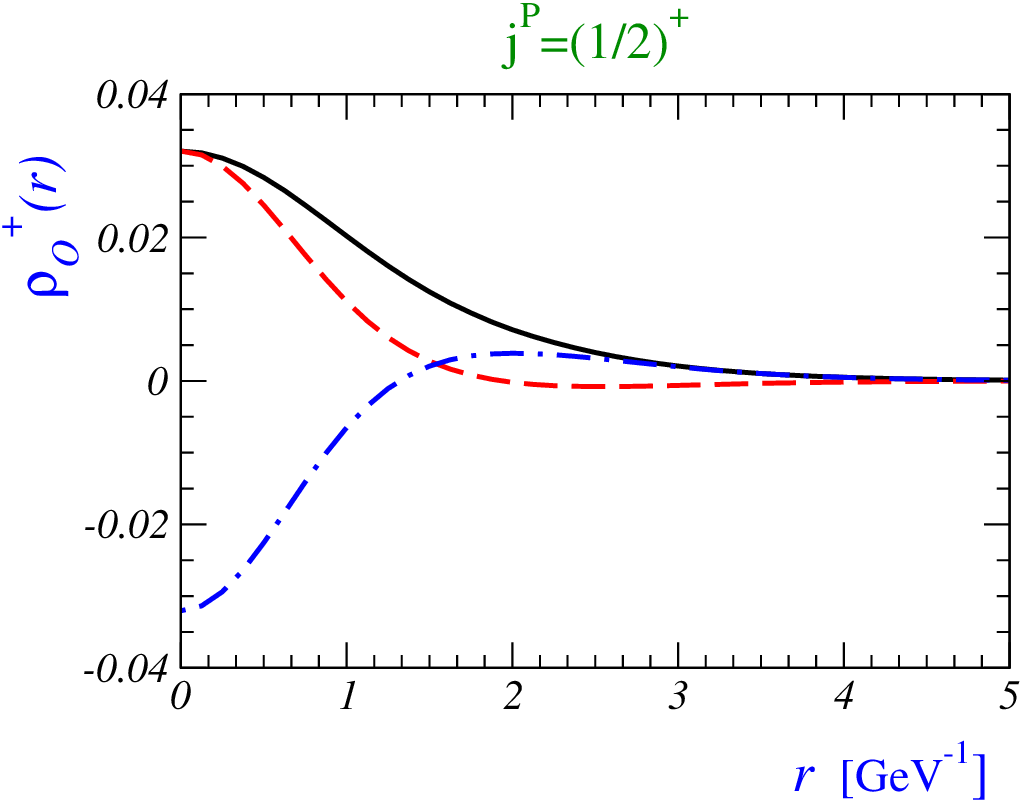}}
\caption{\footnotesize{\sl 
Plots similar to those presented in fig.~\ref{fig:1} but obtained by using the Bakamjian--Thomas type quark model, discussed in the text. }} 
\end{center}
\end{figure}
\bea
\label{28e}
g_S^+ \simeq 0.30\,, \qquad \qquad \widetilde g\equiv  g_A^+ \simeq 0.13\,.
\eea
On the other hand, for the mean square radii, defined in (\ref{16be}), we get
\bea
\label{16ce}
\langle r^2\rangle_V^+\ = 5.42\ \gev^{-2}, \ \ \langle r^2\rangle_S^+\ = 2.73\ \gev^{-2},  \ \ \langle r^2\rangle_A^+\ = -\ 0.015\ \gev^{-2}\,,
\eea
that, of course, satisfy a relation of the type (\ref{25ae}). All three densities obtained by using the BT model are compared with the lattice QCD results in Fig.~\ref{fig:2}.

\section{Comparison of the predictions of the quark models with the lattice  spatial distributions} \label{comparaison-reseau}

\subsection{Short description of the lattice results}

In this section we briefly remind the reader of several details concerning the recent results for the radial distributions discussed in this paper as obtained from the QCD simulations on the lattice. 
For more details, please see ref.~\cite{emmanuel}. The calculation of spatial distributions is performed with a static heavy quark source, and with the light Wilson improved quark action. The gauge field configurations, containing $N_F=2$ mass degenerate light quark flavors, were obtained by using the Iwasaki-type of the lattice Yang-Mills action~\cite{CP-PACS}. The bare coupling constant $g_0$ is chosen according to  $\beta=6/g_0^2=2.1$, and the associated value of the lattice spacing is found to be around $a^{-1} \simeq 2~\gev$. Although dynamical quarks are considered, no attempt has been made to include disconnected diagrams. To improve the signal--to--noise ratio a suitable hyperblocking of the heavy quark propagator (Wilson line) has been implemented. The lattice size used in that computation, $24^3\times  48$, means that the spatial extension of the lattice is $12~$GeV$^{-1}$, so that due to periodic boundary conditions, the points up to radial distances $12~a \simeq 6~$GeV$^{-1}$ could actually be explored. Also, all directions for the spatial points of the lattice have been exploited, allowing for a large averaging and giving a very dense set of points except at small distance.  The four values of  the bare light quark mass (each distinguished by its Wilson parameter $\kappa_q$) have been chosen around the strange quark mass.~\footnote{ More specifically
 $\kappa_q\in \{ 0.1357, 0.1367, 0.1374, 0.1382\}$, with $\kappa_q=0.1374$ being the closest to the strange quark mass. The critical $\kappa_c=0.138984(13)$ corresponds to the chiral limit~\cite{CP-PACS}.} Moreover the valence and the sea-quark masses have been kept degenerate, i.e. the quarks in loops have consistently the same mass as the valence ones. The physical densities are obtained from the ones computed on the lattice 
 via  
\bea
\rho_{O}(r)=1/a^3~Z_{O} \, \bar \rho_{O}^{\rm latt}(r/a)\,, \label{lattice-phys}
\eea 
where $a$ is the lattice unit (lattice spacing), $\bar \rho_{O}^{\rm latt }(r/a)$ stand for the bare lattice data with $O\in{V,S,A}$, and $Z_O(g_0^2)$ is a relevant renormalization constant, all being determined in the same lattice calculations. 
The renormalization constants $Z_{V,A}(g_0^2)$ have been calculated nonperturbatively through Ward identities, and for $\kappa_q=0.1374$, their ${\cal O}(a)$-improved values are $Z_V=0.814$, $Z_A=0.847$. Since these values were found to be consistent with what one obtains by using the boosted perturbative estimates, we use the perturbatively evaluated $Z_S^{\msbar}(2\ \gev)=0.639$. Notice that the above renormalization constant $Z_V$ ensures the vector current  conservation in the lattice regularization scheme, i.e.
\bea
\int \rho_V(r) \, d\vec r = 1\,.
\eea
The corresponding integral concerning the axial current differs from $1$, it is often denoted as  $\widehat g$ or $g_A$, and has a physical meaning. It is related to the pionic coupling $g_{B^\ast B\pi}$ or $g_{D^\ast D\pi}$, up to power corrections.  The latter is a physically observable quantity that can be measured through the strong decay of $D^* \to D \pi$, and has been discussed at length for many years.~\footnote{See a discussion in ref. \cite{dirac1999}.} It also controls the magnitude of loop effects in chiral perturbation theory for heavy-light mesons~\cite{casalbuoni}.

These densities computed on the lattice are presented together with the quark model results in Figs.~\ref{figGROUND},\ref{figEXC}, and in Tables~\ref{tab1},\ref{tab2},\ref{tab3}.

\subsection{Difficulties in the comparison between quark models and lattice QCD} \label{introductory}

\subsubsection{General remarks}

One is faced with the problem of comparing field theory with a naive quantum mechanical model. 
These are completely different approaches, and it could not be expected that they lead to so similar results. 
Agreement is, however, possible probably because the quark models take efficiently into account nonperturbative physical features, 
like the confinement, or the spontaneous generation of quark constituent masses. Those features make QCD quite different from 
what one imagines a field theory to be. Indeed field theory is often envisaged through the familiar, but inadequate, perturbative picture. 
However, those nonperturbative features can be captured by quark models only in an intuitive way, or in average, in an approximate picture. 
As for perturbative effects, they too can be included in quark models (e.g. through radiative corrections, see \cite{gavela}). 
It must not be forgotten that part of all those effects may happen to be taken into account by the adjustment of parameters of the quark models 
on the experimental data.
On the other hand lattice QCD (LQCD) is only an approximation to QCD, with its own errors, and with parameters adjusted to reproduce some experimental data. 
Yet, the essential difference with respect to quark models is that  LQCD is a systematic approach based on field theory, allowing one to approach QCD
as much as one would like by increasing the physical volume of the lattices, the bare coupling constant, and including all the physical quark flavors in the background gauge field configurations. \\

{\bf Remark on the scalar density}: Physically, the scalar density multiplied by a light quark mass perturbation results in a small perturbation of the energy levels of the heavy-light state, i.e. 
\bea
\delta E = \delta m_q \times g_S\,.
\eea
In contrast to vector and axial density the scalar one corresponds to a divergent operator so that its renormalization constant, $Z_S$, is scheme and scale dependent. 
Since the mass renormalization constant cancels against the one related to the scalar charge, there is no need to specify the scheme or scale in the above equation. 
However, for our purpose where we consider the scalar density itself the problem of renormalization should be addressed. The renormalized scalar density computed on the lattice has no real intrinsic physical meaning, which makes problematic a comparison between the quark models and lattice results in this case.  For that reason it is more judicious to divide the scalar density by the scalar charge which is renormalized by the same constant $Z_S$, and therefore the scale and scheme dependence cancels out in the ratio. In other words, we expect  
\bea\label{eq:44}
{\rho_{\rm QM}(r)\over g_S^{\rm QM} }\approx  {\rho_{\rm latt}(r)\over g_S^{\rm latt}} \,.
\eea  \\

 {\bf Disconnected diagrams}: In the field theory calculation, with $N_F \neq 0$, there exist contributions coming from the insertion of the density operator on an isolated quark loop. These are called {\em disconnected diagrams}, meaning that the loop
is not connected to the valence quark lines except by the gluon field. While they vanish automatically for the isovector operators, they may be important in certain cases with isoscalar operators. Even if they can be sizable, such contributions have been ignored in our lattice calculation of densities~\cite{emmanuel}.  Concerning a comparison with quark models, neglecting the disconnected diagrams is just what we have to do, since of course the quark models do not include these effects in the calculation of currents either. We should simply avoid considering currents implying such contributions, which we are of course free to do. 
The exception is the discussion of the meson mass differences due to differences of quark masses.\\

 {\bf Choice of the light quark mass}: While the heavy quark mass is taken to be infinitely large, a more delicate issue is to decide which light quark mass to consider for comparison between quark model results and those obtained in LQCD. There are in fact two distinct questions:
\begin{itemize}
\item[(i)]  How to fix the mass of the light quark? We should fix it both on the lattice and in the quark models to correspond to the same physical spectrum. In particular,  using $\kappa_q=0.1374$ corresponds to the strange heavy-light meson states. We find that the same states are described with $c_0=-0.1$~GeV in eq.~(\ref{eq8}) [{\it Dirac model}], and $m_s=0.42$~GeV [{\it BT model}].
\item[(ii)]  For which light masses the comparison between lattice and quark model results could be made optimal? It turns out that both lattice QCD and quark models are more predictive when away from the chiral limit. A comparison is better for the strange light quark mass, for which the information on physical spectrum is available too. Therefore we choose to work with strange heavy-light meson states.
\end{itemize}
Concerning the point b), the cost of numerical simulation of QCD on the lattice becomes prohibitively large when going closer to the chiral limit, and reaching the physical $u,d$ quarks necessitates chiral extrapolations. The strange quark mass is, instead, directly accessible from the current lattice studies. On the other hand, the predictions of our quark models can be directly formulated at will for any light quark mass. At present the constituent quark models do not include the effects of chiral loops which might be significant at very small quark masses, such as the case of electric radii of mesons and baryons whose expected logarithmic rise at small pion masses is not satisfied by the models~\cite{Gasser:1984ux}. \\

{\bf Respective ranges of validity of quark models and lattice QCD}:
Some comments are in order regarding the respective domains of validity of quark models and lattice QCD as a function of $r$. 
\begin{itemize}
\item[1)] At short-distances, one would expect hard gluons and quantum loop corrections to play a dominant role. In particular, the interquark potential should be dominated by hard gluon exchange. 
This effect is only mimicked by the Coulomb potential in the quark models, which is an effective potential, with a range much beyond the proper UV region, as we already discussed in sec.~\ref{parameters}. 
In lattice QCD, on the other hand, one should be particularly careful when probing the small distance region where the discretization errors might become large.
\item[2)] At long-distances, the confining potential is expected to be modified by pair creation: the two-body potential should become flat, because the energy is now entirely converted into pairs. 
This flattening is indeed demonstrated in lattice QCD  for $N_F > 0$, around $r=1.25~$fm, which means $r=6.25~$GeV$^{-1}$, for the $Q \bar{Q}$ potential, by studying the mixing between $Q \bar{Q}$ ($b \bar{b}$) and di-meson ($B \bar{B}$) states~\cite{bali}.
 One would then expect the quark model density distributions to have too steep a decrease at large $r$, because the confining potential is too strong there. In practice we observe that a good agreement between quark models and lattice data extends farther than naively expected (up to $4~\gev^{-1}$ and even beyond).
\end{itemize}

Further difficulty might arise from the fact that we use a lattice with a finite size. Moreover, due to the periodic boundary conditions, only half of its distances can be explored, i.e. $r \lesssim 6$ GeV$^{-1}$.  Fortunately all distributions discussed in this paper are fast decreasing functions of $r$ and die out before reaching the boundaries, as shown in ref.~\cite{emmanuel}. The size of the available distances explored in that paper were shown to be enough for this study. In other words, $6$~GeV$^{-1}$ in physical units appears to be already far beyond the mean radius of the ground state $\simeq \sqrt{5}~$GeV$^{-1}$, and beyond the observed range of agreement
of the quark model and lattice distributions (around $4~\gev^{-1}$).

\subsubsection{Errors in quark models} 

In the quark model it is difficult to make a priori statements about errors since it is not a well defined approximation scheme to QCD. The
only real way to gauge the errors is {\em a posteriori}, through comparison with experimental data, or with more rigorous theoretical approaches. 
From comparison with experiment, we know that we can obtain a good description of spectra and of a certain set of matrix elements, but we know that it is not a quantitative statement, that the agreement is very dependent on the observable considered, and on the particular model. No model of course can claim an accurate prediction of all possible observables.

Let us just underline two important facts :

a) Not every quark model shows an extensive agreement with experiment, and in general very few models have been tested on a very large set of data. This is an argument in favor of using the GI model which shows  good agreement with a very large set of hadronic masses and  decay widths. Concerning the decay widths, however, it should be stressed that the basic eigenvalue equation of GI has to be supplemented by additional modeling like the elementary quantum emission model, or the quark pair creation model. For the purpose of calculating spatial distributions, we insert it into our Bakamjian--Thomas approach to current matrix elements, a combination which has been shown to be successful in many places. The success of this approach in predicting important features of Isgur-Wise functions is an important argument: it tests both the potential model and the treatment of hadron motion.  

More generally speaking, one can say that there is growing evidence, after many years, that only relativistic models can provide a consistent picture of processes involving light quarks. 
This applies especially to the internal velocities.

b) As for the spatial distributions of the static heavy-light mesons considered in the present paper, we think a priori that a static limit is a favorable place for quark models; and also favorable is that we consider only elastic matrix elements, without considering transitions involving radial excitations that would be more sensitive to errors. In particular, this is the best place for quark models based on the Dirac equation, while, for instance, they would not be suited for annihilation processes. The potential that we use in the Dirac equation has not been tested on such a large set of data as the GI one, simply because it is specific to heavy-light quark systems. But it appears in practice to be close to the GI one.

A third favorable condition is to choose as light quark the $s$ rather than the $u,d$
ones. This allows, as we said, to avoid as much as possible the phenomenon of spontaneous symmetry breaking, beyond the scope of most constituent quark models.

This favorable context being taken into account, it must be said that the degree of agreement which we shall display below for both types of models, is surprising in the context of quark models, especially for the $(1/2)^+$ state. Quark models are often thought of as capable of reproducing the rudimentary features of hadrons, and here we observe a very good agreement with lattice QCD for detailed information on light quark current densities as functions of the distance from the static heavy quark source.

\subsubsection{Lattice errors}

We must recall the traditional distinction between {statistical} and {systematic} errors. Statistical errors are easy to compute. They are often the only ones quoted in the tables and figures, but they are most often not the only ones, since the systematic errors are difficult to evaluate, and difficult to reduce. 

In the present case, the statistical errors are very small, as it can be seen in Figs.~\ref{figGROUND},\ref{figEXC}. As for the sources of systematic uncertainties, they can be divided in discretization errors and finite volume effects. As we already mentioned, the results obtained in ref.~\cite{emmanuel} are obtained using the lattice box of the size $L=6~\gev^{-1}$, which appears to be larger than the mean radius of the heavy-light meson states discussed in this paper. For that reason, and since we are not working with very light quarks, we believe that systematic errors due to finite volume are negligible. 

As for discretization errors, in ref.~\cite{CP-PACS} it was demonstrated on the case of hadron masses and decay constants that  working with the strange quark mass and with ${\cal O}(a)$-improved Wilson action, combined with the Iwasaki gauge action, a bulk of discretization errors is minimized. It appears that the results for physical quantities computed in ref.~\cite{CP-PACS} at $\beta=2.1$ are very close to their continuum limit, and the difference is always below $10~\%$. We can therefore assume that a similar situation holds true in the present study.  Apart from ``intrinsic" discretization errors, there are also those related to the conversion of results from the lattice to physical units, such is the case in eq.~(\ref{lattice-phys}). For that it is necessary to fix the lattice spacing by comparing one and the same quantity computed on the lattice with its physical value. The most convenient way to fix the lattice spacing is via the Sommer radius $r_0$~\cite{sommer} that can be accurately computed on the lattice, but the drawback is that its physical value is not known. In ref.~\cite{CP-PACS} it was shown that the conventionally adopted $r_0=0.5~{\mathrm fm}$ is compatible with what is obtained if the lattice spacing was fixed by using the $\rho$-meson mass. Note, however, that the determination of lattice spacing is immaterial when computing the adimensional couplings  $g_{V,A,S}^{(+)}$.

Finally, we should comment on the sea-quark mass dependence. In this paper we make a comparison between the distributions obtained within two classes of quark models and those obtained from the QCD simulations on the lattice with both the valence and the sea quarks close to that of the physical strange quark. A more physical  comparison would require an extrapolation in the sea quark to the physical light ``$u/d$" quark mass. We tested that changing that mass to half of the strange quark mass (corresponding to $\kappa_{\rm sea}=0.1382$) does not change our distributions at all. Whether or not that remains the same when the sea-quark mass becomes much lighter is not clear. The CP-PACS Collaboration, by working with the same Iwasaki gauge action, checked that the masses and decay constants for both light-light~\cite{CP-PACS} and  for heavy-light mesons~\cite{CP-PACShl} do not depend on the variation of the sea-quark mass. A similar conclusion for the heavy-light mesons has been reached by using the actions allowing one to probe very light sea quark masses, namely, the improved staggered quarks~\cite{Davies:2010ip} and the twisted mass QCD~\cite{etmc-d}.  We will therefore assume that a dependence on the sea-quark mass is small. 

For the purpose of comparison with quark models, in the following we will assume that the systematic errors of the lattice results are small, at the $10\%$ level, and dominated by the discretization effects. The conclusions of the present paper should be understood to be valid within the lattice systematics.

\subsection{Quantitative comparison of the two quark models and of lattice QCD}
\label{comparaison}

\begin{figure}[h!]
\begin{center}
\epsfig{file=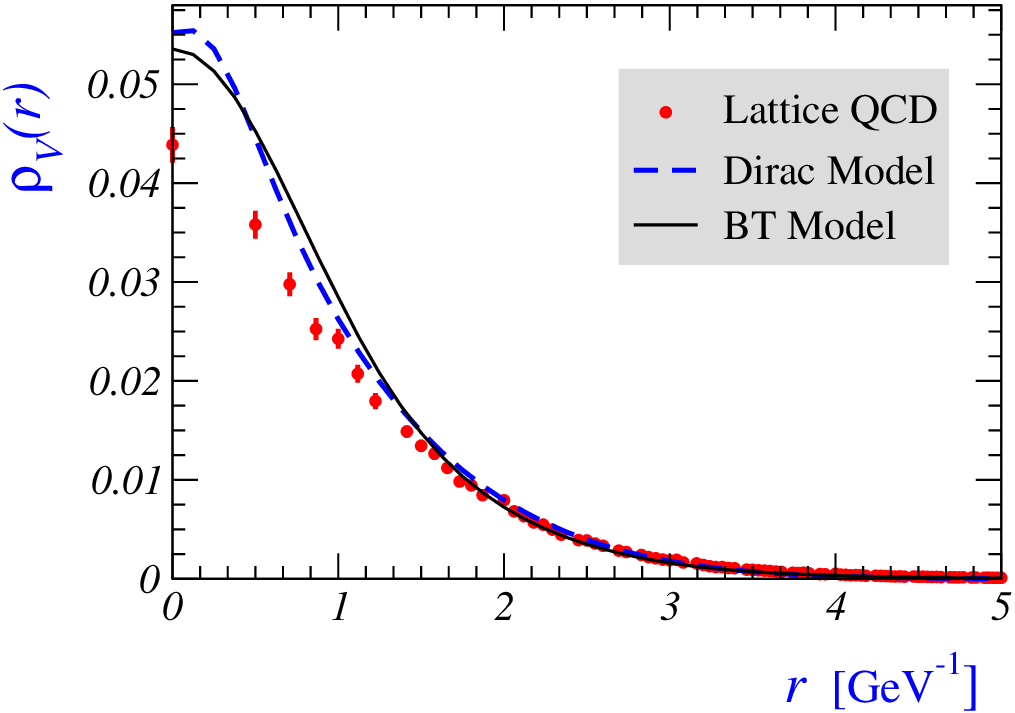, height=6cm}\\ 
{\phantom{\Huge{l}}}\raisebox{-.1cm}{\phantom{\Huge{j}}}
\epsfig{file=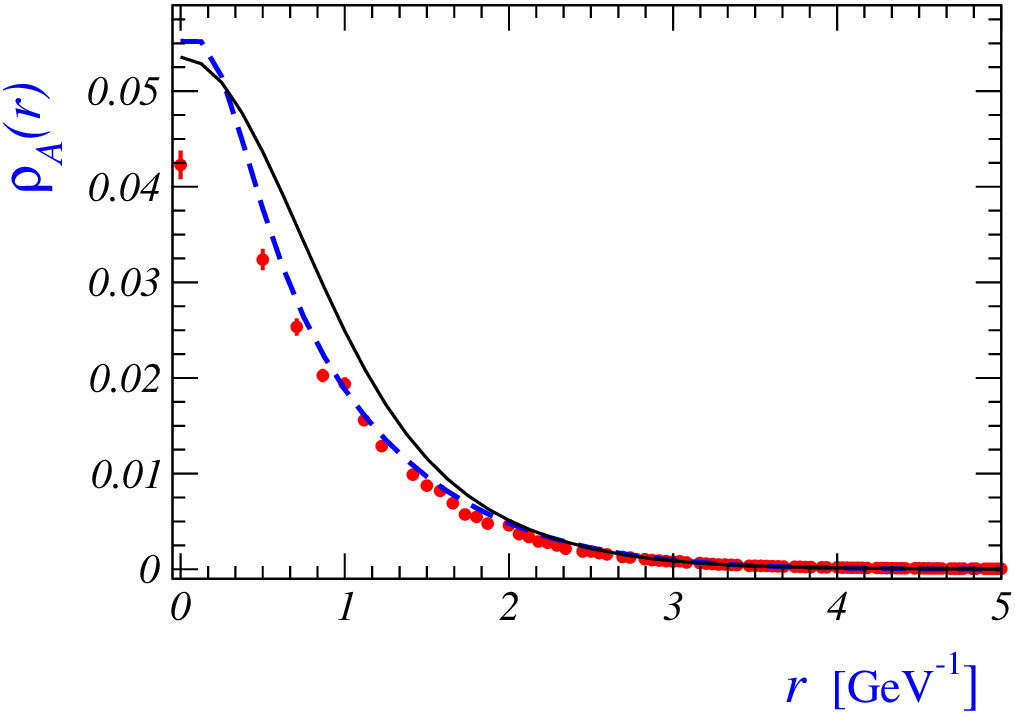, height=6cm}\\
{\phantom{\Huge{l}}}\raisebox{-.1cm}{\phantom{\Huge{j}}}
\epsfig{file=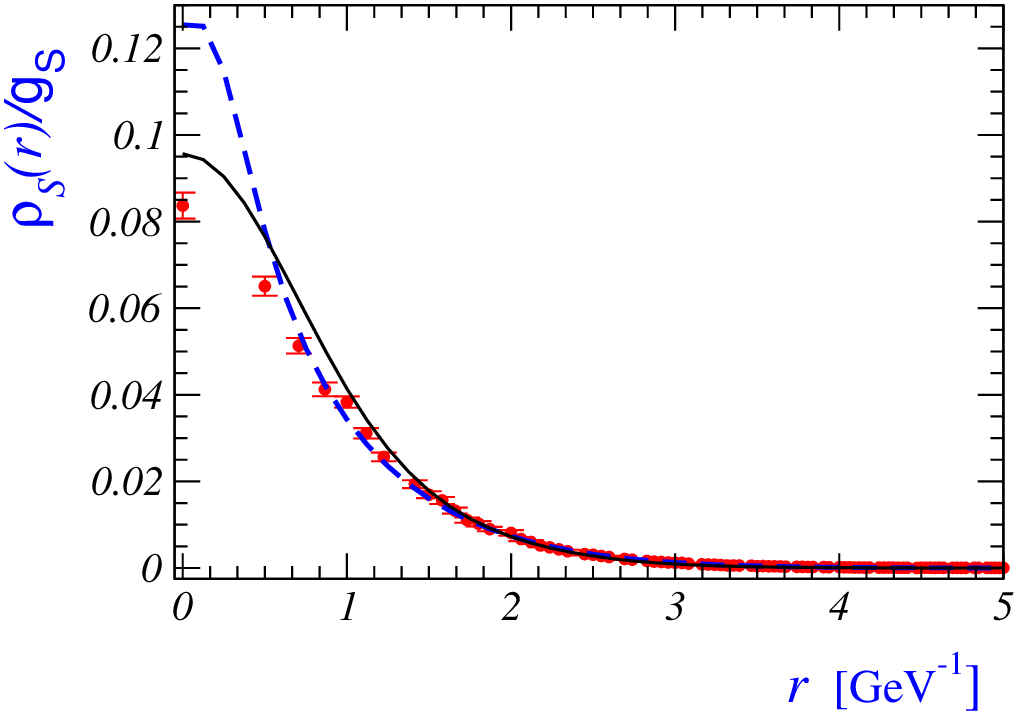, height=6cm}
\caption{\label{figGROUND}\footnotesize{\sl 
Comparison of the results for three densities [vector $\rho_V(r)$, axial $\rho_A(r)$, and scalar $\rho_S(r)/g_S$] for the lowest doublet of the static heavy-light mesons, $j^P=(1/2)^-$, as obtained by using two quark models discussed in the text, namely a Dirac model (dashed line) and a model \`a la Bakamjan-Thomas (solid line). They are displayed  together with results obtained in lattice QCD reported in ref.~\cite{emmanuel}. }} 
\end{center}
\end{figure}

\begin{figure}[h!]
\begin{center}
\epsfig{file=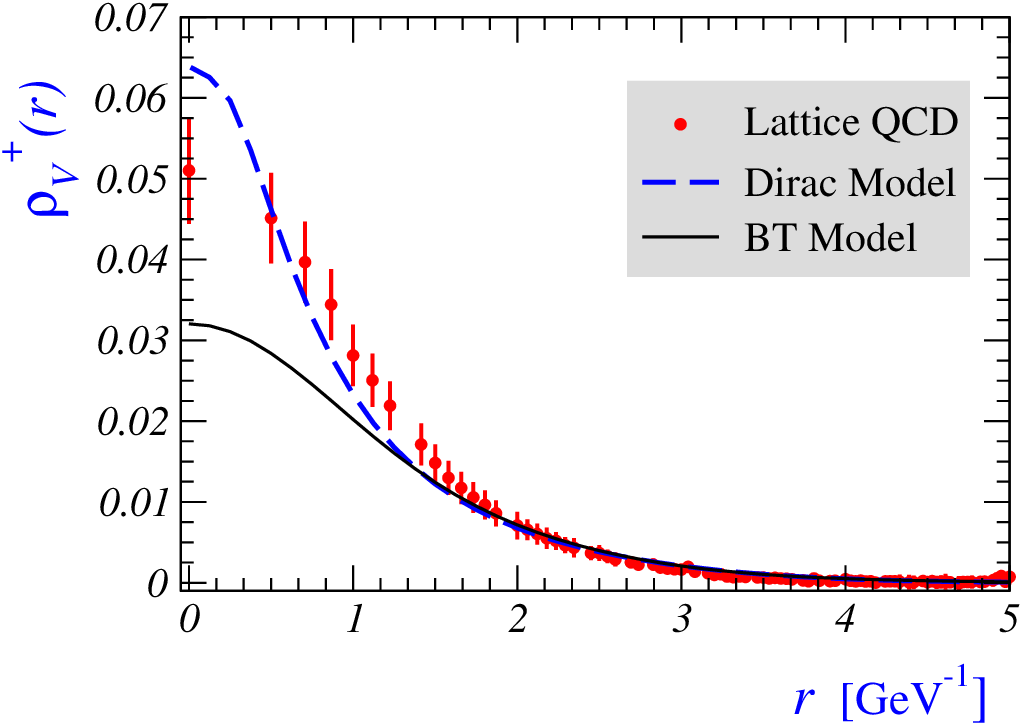, height=6cm}\\ 
{\phantom{\Huge{l}}}\raisebox{-.1cm}{\phantom{\Huge{j}}}
\epsfig{file=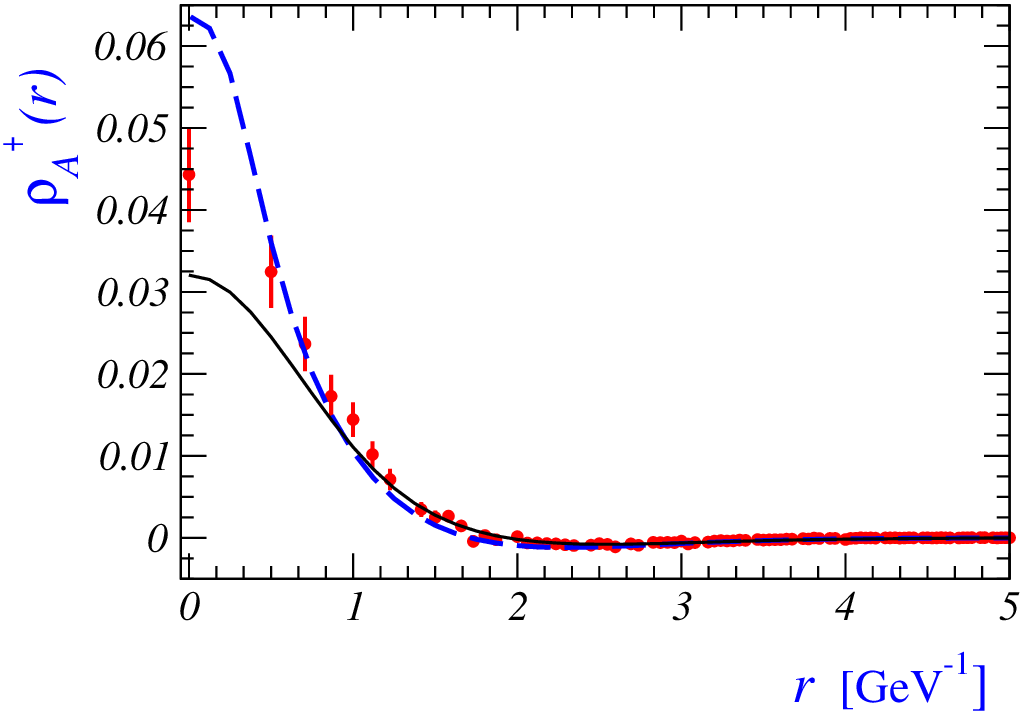, height=6cm}\\
{\phantom{\Huge{l}}}\raisebox{-.1cm}{\phantom{\Huge{j}}}
\epsfig{file=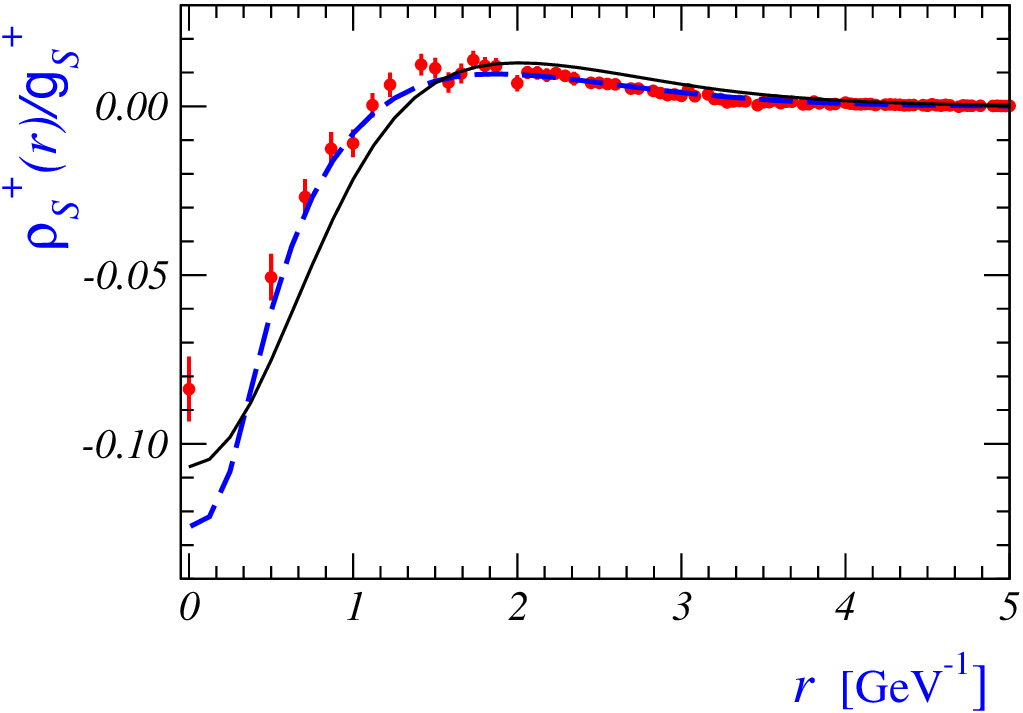, height=6cm}
\caption{\label{figEXC}\footnotesize{\sl 
The same as in fig.~\ref{figGROUND} but for the orbitally excited doublet, i.e. $j^P=(1/2)^+$. }} 
\end{center}
\end{figure}

\begin{figure}[h!]
\begin{center}
\epsfig{file=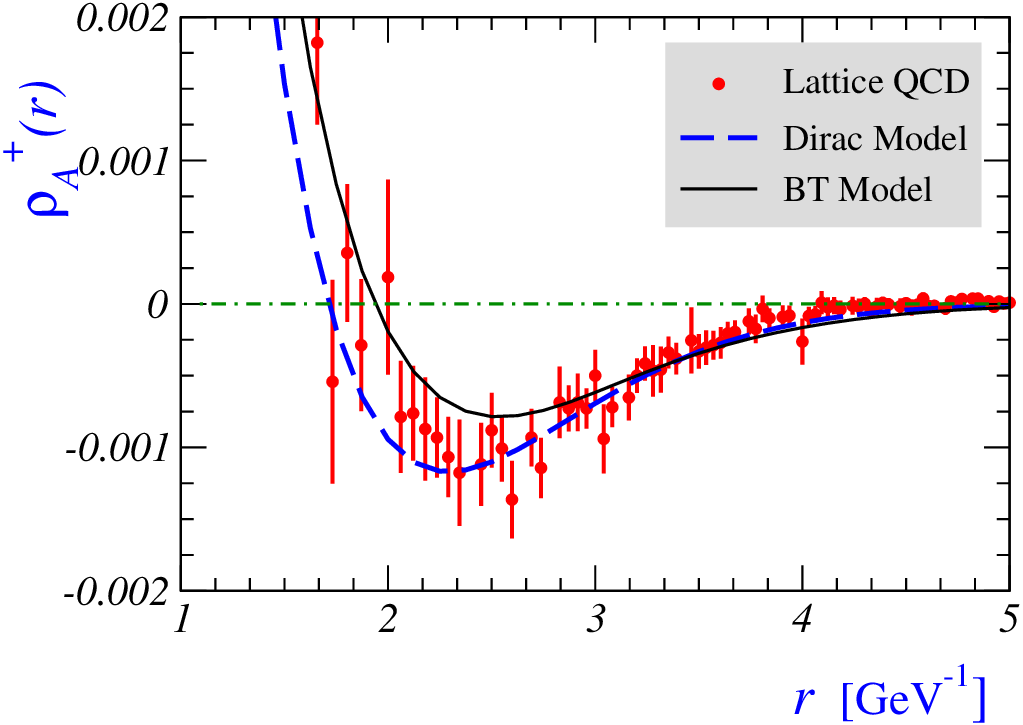, height=8cm}\\
{\phantom{\Huge{l}}}\raisebox{-.1cm}{\phantom{\Huge{j}}}
\epsfig{file=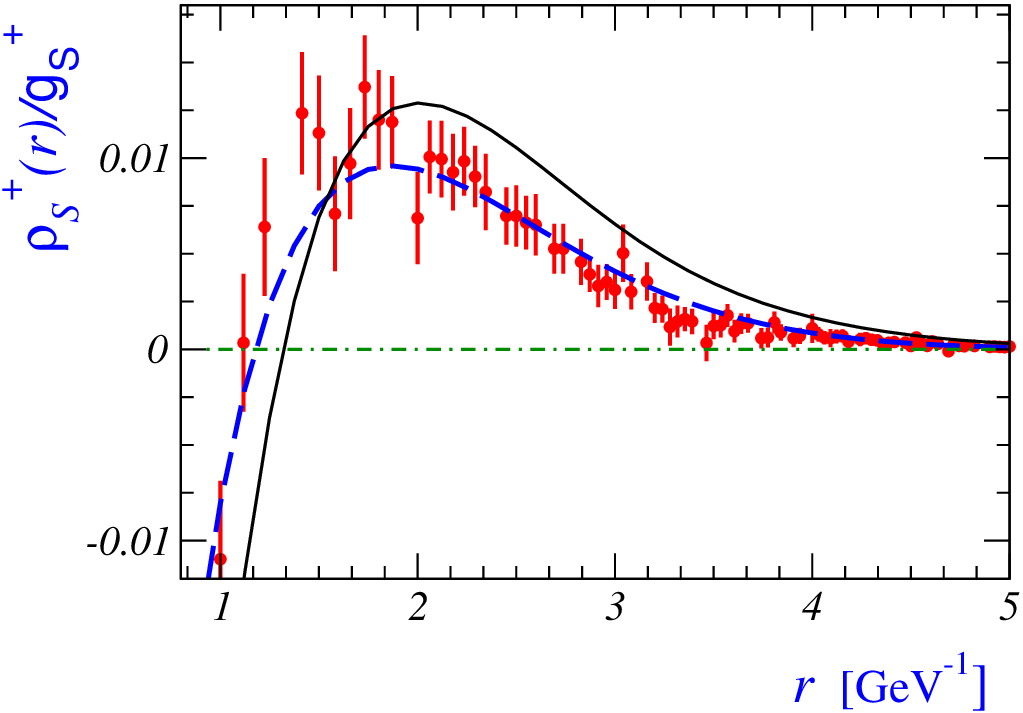, height=8cm}
\caption{\label{figZOOM}\footnotesize{\sl 
Plots showing the region in which the axial and scalar  densities in fig.~\ref{figEXC} change the sign, $j^P=(1/2)^+$. Dot-dashed horizontal line, corresponding to zero densities, is added for the reader's convenience. }} 
\end{center}
\end{figure}

Results of the two quark models, and those obtained on the lattice are listed in Tables~\ref{tab1}, \ref{tab2}, \ref{tab3}, and displayed in Figs.~\ref{figGROUND}, \ref{figEXC}, \ref{figZOOM}. Following the discussion around eq.~(\ref{eq:44}) in the case of scalar density we compare the ratios $\rho_S^{(+)}(r)/g_S^{(+)}$.

\begin{table}[hbt]
{\scalebox{.94}{\begin{tabular}{|c||c|c|c||c|c|c|}
\cline{2-7}
\multicolumn{1}{c||}{}&\multicolumn{3}{|c||}{$\displaystyle{j^P=(1/2)^{-^{}}}$}&\multicolumn{3}{|c|}{$j^P=(1/2)^+$}\\
\hline
{\phantom{\Huge{l}}}\raisebox{-.1cm}{\phantom{\Huge{j}}}
$r\ [\gev^{-1}]$ & Dirac & BT & lattice QCD & Dirac  & BT & Lattice QCD\\
\hline\hline
{\phantom{\Huge{l}}}\raisebox{-.1cm}{\phantom{\Huge{j}}}
0. & 0.0534  &  0.0536 & 0.0446(12) & 0.064 & 0.0320 & 0.0522(44)\\  
\hline
{\phantom{\Huge{l}}}\raisebox{-.1cm}{\phantom{\Huge{j}}}
0.5 & 0.0445 & 0.0453& 0.0345(8) & 0.0462 & 0.0284 & 0.0427(34)\\
\hline
{\phantom{\Huge{l}}}\raisebox{-.1cm}{\phantom{\Huge{j}}}
1.& 0.0262 & 0.0285 & 0.0229(6) & 0.0233 & 0.0202 & 0.0275(24)\\
\hline
{\phantom{\Huge{l}}}\raisebox{-.1cm}{\phantom{\Huge{j}}}
1.5 &  0.0149 & 0.0148 & 0.0135(4) & 0.0122 & 0.0124 & 0.0149(15)\\
\hline
{\phantom{\Huge{l}}}\raisebox{-.1cm}{\phantom{\Huge{j}}}
2.  &  0.00770 & 0.00724 & 0.00733(24) & 0.00678 & 0.00715 & 0.0072(11)\\
\hline
{\phantom{\Huge{l}}}\raisebox{-.1cm}{\phantom{\Huge{j}}}
2.5 &   0.00394 & 0.00349 & 0.00375(13) & 0.00383 & 0.00397 & 0.00328(51)\\
\hline
{\phantom{\Huge{l}}}\raisebox{-.1cm}{\phantom{\Huge{j}}}
3.&  0.00178 & 0.00161 & 0.00185(7) & 0.00208 & 0.00208 & 0.00145(28)\\
\hline
{\phantom{\Huge{l}}}\raisebox{-.1cm}{\phantom{\Huge{j}}}
3.5 & 0.000724 & 0.000724 & 0.000895(38)  & 0.00105 & 0.00104  & 0.00065(20)\\
\hline 
{\phantom{\Huge{l}}}\raisebox{-.1cm}{\phantom{\Huge{j}}}
4. & 0.000265 & 0.000332 & 0.000431(18) & 0.000476  & 0.000507 & 0.00030(11)\\
\hline
{\phantom{\Huge{l}}}\raisebox{-.1cm}{\phantom{\Huge{j}}}
5. &  0.0000258 & 0.0000665 & 0.000102(5) & 0.0000683 & 0.00011 & 0.000076(43)\\
\hline
\end{tabular} 
}}
\caption{\footnotesize{\sl Comparison between the two models and lattice QCD for  $\rho_V$ and  $\rho_V^+$ in physical units $[\gev^{3}]$}}
\label{tab1}
\end{table} 

\vskip 0.5cm
\begin{table}[hbt]
{\scalebox{.92}{\begin{tabular}{|c||c|c|c||c|c|c|}
\cline{2-7}
\multicolumn{1}{c||}{}&\multicolumn{3}{|c||}{$\displaystyle{j^P=(1/2)^{-^{}}}$}&\multicolumn{3}{|c|}{$j^P=(1/2)^+$}\\
\hline 
{\phantom{\Huge{l}}}\raisebox{-.1cm}{\phantom{\Huge{j}}}
$r\ [\gev^{-1}]$ & Dirac  & BT & lattice QCD & Dirac &  BT & Lattice QCD \\
\hline\hline
{\phantom{\Huge{l}}}\raisebox{-.1cm}{\phantom{\Huge{j}}}
0. & 0.0530  & 0.0536 & 0.0436(10)  & 0.0640 & 0.0320 & 0.0445(39)\\
\hline
{\phantom{\Huge{l}}}\raisebox{-.1cm}{\phantom{\Huge{j}}}
0.5 & 0.0377  & 0.0437 & 0.0313(7) & 0.0360  & 0.0245 & 0.0320(29)\\
\hline
{\phantom{\Huge{l}}}\raisebox{-.1cm}{\phantom{\Huge{j}}}
1.& 0.0188 & 0.0250 & 0.0180(4) & 0.0105 & 0.0110 & 0.0135(13)\\
\hline
{\phantom{\Huge{l}}}\raisebox{-.1cm}{\phantom{\Huge{j}}}
1.5 & 0.00969   & 0.0116 & 0.00890(21) & 0.00153  & 0.00276 & 0.00273(63)\\
\hline
{\phantom{\Huge{l}}}\raisebox{-.1cm}{\phantom{\Huge{j}}}
2.  & 0.00481  & 0.00512 & 0.00406(17) & -0.000943  & -0.000192 & -0.00046(62)\\
\hline
{\phantom{\Huge{l}}}\raisebox{-.1cm}{\phantom{\Huge{j}}}
2.5 & 0.00224   & 0.00219 & 0.00179(8) & -0.00110  & -0.000786 & -0.00087(22) \\
\hline
{\phantom{\Huge{l}}}\raisebox{-.1cm}{\phantom{\Huge{j}}}
3.& 0.000965  & 0.000877 & 0.00080(4) & -0.000693   &  -0.000616 & -0.00128(24) \\
\hline
{\phantom{\Huge{l}}}\raisebox{-.1cm}{\phantom{\Huge{j}}}
3.5 &0.000377  & 0.000346 & 0.00037(2) & -0.000333  & -0.000344  & -0.00167(51)\\
\hline 
{\phantom{\Huge{l}}}\raisebox{-.1cm}{\phantom{\Huge{j}}}
4. & 0.000133 & 0.000140 & 0.000182(8) & -0.000131   & -0.000164 & -0.00142(73)\\
\hline
{\phantom{\Huge{l}}}\raisebox{-.1cm}{\phantom{\Huge{j}}}
5. & 0.0000123  & 0.0000199 & 0.0000525(28) &  -0.0000119  & -0.0000250 &-- \\
\hline
\end{tabular}
}}
\caption{\footnotesize{\sl Comparison between the two models and lattice QCD for  $\rho_A$ and  $\rho_A^+$ in physical units $[\gev^{3}]$}}
\label{tab2}
\end{table} 
\vskip 0.5cm
\begin{table}[hbt]
{\scalebox{.92}{\begin{tabular}{|c||c|c|c||c|c|c|}
\cline{2-7}
\multicolumn{1}{c||}{}&\multicolumn{3}{|c||}{$\displaystyle{j^P=(1/2)^{-^{}}}$}&\multicolumn{3}{|c|}{$j^P=(1/2)^+$}\\
\hline
{\phantom{\Huge{l}}}\raisebox{-.1cm}{\phantom{\Huge{j}}}
$r\ [\gev^{-1}]$ & Dirac  & BT & lattice QCD & Dirac &  BT & Lattice QCD \\
\hline\hline
{\phantom{\Huge{l}}}\raisebox{-.1cm}{\phantom{\Huge{j}}}
0. & 0.1200  & 0.0957 & 0.0854(25)  & -0.1251 &-0.1067 & -0.0739(97)\\
\hline
{\phantom{\Huge{l}}}\raisebox{-.1cm}{\phantom{\Huge{j}}}
0.5 & 0.0780 & 0. 0764 & 0.0635(18)  &-0.0608  &-0.0753 & -0.0444(69)\\
\hline
{\phantom{\Huge{l}}}\raisebox{-.1cm}{\phantom{\Huge{j}}}
1.& 0.0341  &  0. 0414 & 0.0363(11) &-0.00804  &-0.0217  & -0.0080(34)\\
\hline
{\phantom{\Huge{l}}}\raisebox{-.1cm}{\phantom{\Huge{j}}}
1.5 & 0.0161  &  0.0179 & 0.0174(7)  & 0.00745 &0.0069  &0.0090(27) \\
\hline
{\phantom{\Huge{l}}}\raisebox{-.1cm}{\phantom{\Huge{j}}}
2.  & 0.0073  & 0.0072 & 0.00742(48)  & 0.00941&0.0128  & 0.0109(23)\\
\hline
{\phantom{\Huge{l}}}\raisebox{-.1cm}{\phantom{\Huge{j}}}
2.5 & 0.0032   &0.0028  & 0.00298(25) &0.00698  &0.0105 & 0.0073(19)  \\
\hline
{\phantom{\Huge{l}}}\raisebox{-.1cm}{\phantom{\Huge{j}}}
3.& 0.0013  & 0.00091  & 0.00117(12)  &0.00408  &0.0065  & 0.0035(11) \\
\hline
{\phantom{\Huge{l}}}\raisebox{-.1cm}{\phantom{\Huge{j}}}
3.5 & 0.00045   & 0.00028  & 0.000456(71)  &0.00200  &0.0034  & 0.0015(10)  \\
\hline 
{\phantom{\Huge{l}}}\raisebox{-.1cm}{\phantom{\Huge{j}}}
4. &0.00015   &0.000080  & 0.000177(33)  &0.000314  &0.0017 & 0.00054(30) \\
\hline
{\phantom{\Huge{l}}}\raisebox{-.1cm}{\phantom{\Huge{j}}}
5. & 0.000018  &0.000006 & 0.0000246(95)   &0.000102  &0.00031  & 0.000050(54)\\
\hline
\end{tabular}
}}
\caption{\footnotesize{\sl Comparison between the two models and lattice QCD for  $\rho_S/g_S$ and  $\rho_S^+/g_S^+$ in physical units $[\gev^{3}]$}}
\label{tab3}
\end{table}

%\noindent 

The first striking fact is the similarity of the two quark models for the $L=0,~j=1/2$ state, which was rather unexpected considering that they have a very different formal structure. 
On the other hand, the results concerning the $L=1,~j=1/2$ agree less in the region of very low values of $r$, but the overall qualitative agreement is still there. Moreover, the agreement at intermediate and larger values of $r$ is very good. 
We also note that contrary to nonrelativistic models, the axial and scalar densities of the $L=1$ state exhibit nodes in both models, as well as in QCD. 

It is obvious from the tables and plots that the two quark models agree very well with results of lattice QCD. More specifically:
\begin{itemize}
\item  Both models are able to reproduce quantitatively the three lattice distributions on a large range of distances for both $L=0$ and $L=1$ states, i.e.  $1~\gev^{-1}\lesssim r \lesssim 4~\gev^{-1}$. Let us recall that there is no adjustment of parameters of the quark models on these distributions. {\underline{ \sf The parameters were entirely fixed from the fit to the hadronic spectrum}}, as discussed above. We note however that  for the $L=1$ states,  dis\-tri\-bu\-tions predicted by the models considered in this paper somewhat differ between themselves and from the lattice data in the region $r \lesssim 1~\gev^{-1}$. For the ground state, instead, the agreement is quantitatively good everywhere, i.e. including the small $r$-region.

\item Both models reproduce remarkable features of the distributions of excited $L=1,~j=1/2$ states.

a) From Figs.~\ref{figGROUND} and \ref{figEXC} we observe the similarity of the vector charge distributions for $L=0$ and $L=1$ states as obtained by both models and from our lattice data. 
Even though the overall charge is conserved (fixed to $1$), the fact that their distributions are very similar was not a priori obvious.  

b) On the contrary, the axial and scalar distributions for $L=1$ states are very different from the ones  obtained for $L=0$ states. 
In particular, a change of sign is observed for the excited states, a feature not present in the nonrelativistic wave functions and distributions. 
Moreover, {\sl the positions of zeros are quantitatively predicted by the quark models}: in the Dirac [BT] model, it is $r=1.7~$GeV$^{-1}$ [$r=1.937~$GeV$^{-1}$] for the axial density, to be compared with $r=1.85~$GeV$^{-1}$ as obtained on the lattice. The zero  is notably lower for the scalar density, namely in the Dirac (BT) model, we find it to be  $r \simeq 1.2~$GeV$^{-1}$  ($r=1.319~$GeV$^{-1}$), once again in a very good agreement with the lattice QCD result, $r \simeq 1.1~$GeV$^{-1}$.
Furthermore, from Fig.~\ref{figZOOM} we observe that the signs predicted by our models agree with lattice QCD:  in the inner region where the wave function is large, the axial (scalar) distribution is positive (negative), which is the opposite of what a nonrelativistic calculation would give.
\end{itemize}

\subsubsection{Explanation of signs and zeroes. Inner versus outer regions}

To explain how these sign changes occur we consider the simpler Dirac formalism. The three densities are expressed in terms of upper ($f$) and lower ($g$) components of the radial wave functions, as given in eqs.~(\ref{dirac-distr-0}) and (\ref{dirac-distr-1}),  for $L=0$ and $L=1$ states, respectively (both being $j=1/2$).
{\underline{In the nonrelativistic limit}},  $|g_{1/2}^{\mp 1}(r)|^2 \ll |f_{1/2}^{\mp 1}(r)|^2$, which leads to
\bea  
&&\qquad~\qquad\rho_V(r)= \rho_S(r) =\rho_A(r) = |f_{1/2}^{- 1}(r)|^2,  \nonumber \\
&&\\
&& \rho^+_V(r)= \rho^+_S(r)=|f_{1/2}^{ 1}(r)|^2 ,\qquad\quad \rho^+_A(r) = -{1\over 3}|f_{1/2}^{ 1}(r)|^2\,.\nn
\eea
$\rho_V(r)$ and $\rho^+_V(r)$ are of course positive and,  when integrated over the three spatial dimensions,  should give unity. In the nonrelativistic model the same appears to hold true for the scalar densities. 
The axial densities differ instead in sign (positive for the ground states, and negative for the excited ones) and the associated charges also differ in their absolute values: for the ground states it is again equal to one, whereas for the excited ones it gives $-1/3$.
The sign of distributions for the ground states remains as such after the relativistic corrections are included because $|f_{1/2}^{-1}(r)|^2$ remains larger than $|g_{1/2}^{- 1}(r)|^2$ for all values of $r$. For the excited states, instead, the relativistic corrections drastically  modify the form of axial and scalar distributions.
In the {\underline{inner}} region, where the wave functions are most important, the relativistic corrections reverse the situation and the upper component $|f_{1/2}^{ 1}(r)|$ becomes smaller than $|g_{1/2}^{ 1}(r)|$. Therefore the axial density is positive instead of negative, and the scalar density is negative instead of positive.
For small $r$ we note that this  dominance of the lower components is in agreement with the expected behavior, $f_{1/2}^{1}(r) \propto r$, $g_{1/2}^{1}(r)(r) \propto r^0$~\cite{landau}. For growing $r$, the upper component $|f_{1/2}^{1}(r)|^2$ wins rather rapidly over $|g_{1/2}^{ 1}(r)|^2$, so that  $|g_{1/2}^{1}(r)|^2 - 1/3 |f_{1/2}^{1}(r)|^2$, and $|f_{1/2}^{1}(r)|^2 -  |g_{1/2}^{1}(r)|^2$, go through zero toward the nonrelativistic sign. 

The fact that the zero of the distributions $\rho^{0^+}_{S}(r)$ is closer to the origin than in the case of  $\rho^{0^+}_{A}(r)$  can also be explained: the coefficient of the upper component is larger in the scalar case, so the return to the nonrelativistic sign happens earlier.

\subsubsection{A $\gamma_5$-symmetry? The true and the false}

In the Dirac model, in the inner region we observe an exchange of magnitudes between the upper and lower components of the radial wave functions in the excited state, with respect to the situation in the ground state. Roughly speaking, 
\bea
|f_{1/2}^{1}(r)|^2 \simeq |g_{1/2}^{-1}(r)|^2,\qquad  |g_{1/2}^{1}(r)|^2 \simeq |f_{1/2}^{-1}(r)|^2\,.
\eea
%\section{}
In fact, there is an approximate $\gamma_5$-symmetry, which can be (with suitable normalization) written as:
\bea
\psi_{(1/2)^+} \simeq \gamma_5\, \psi_{(1/2)^-}\,.
\eea
This symmetry is sufficiently good to explain that the vector and axial distributions of $(1/2)^-$ and $(1/2)^+$ states are similar in this inner region. 
Lower components give a large nonzero value at the origin for all three densities, with the relation dictated by $\gamma_5$ symmetry and the vanishing of $f_{1/2}^{1}(r) \to 0 $ at the origin,
\bea
\rho_V^{+}(0)=\rho_A^{+}(0)=-\rho_S^{+}(0)\,.
\eea
These relations, however, are only valid inside what we call the inner region. 
The situation changes at larger $r$ (outer region) where the $\gamma_5$-symmetry does not hold. This may explain the failure of the prediction made in ref.~\cite{bardeen} concerning the axial couplings, which is solely based on chiral symmetry (see below).

Moments of  $\rho_A$ and $\rho_S$, instead, receive sizable contribution from  the large $r$ region (outer region), where the sign is opposite to the  low $r$ region. 
The consequence is that for axial and scalar density moments,  the approximate $\gamma_5$-symmetry is completely broken, as noticed first in ref.~\cite{damir0406296} for the integral of the axial charge.

The fact that all these highly non trivial features, predicted in a simple manner by the relativistic quark models, are verified by lattice QCD seems a big success for quark models.

We stress the peculiarity of  $L=1$ states at not too large $r$, where the ratio between the upper and lower components of Dirac spinors is reversed with respect to the nonrelativistic limit. Relativity is no longer a correction; it leads to a qualitative change.

Within the Dirac formalism, we can even interpret this difference between an inner and an outer region.
The $\gamma_5$-symmetry corresponds to a vectorlike Hamiltonian. The Dirac kinetic energy and the ``short-distance" Coulomb-like potential are vectorlike, while the sum of the confining potential and the mass term is scalar,
\bea
S(r)=a r + c + m\,.
\eea  
Note, however, that  the ``short" distance is actually not so short, since the strength of the Coulomb-like term is rather large, $\kappa=0.75$. On the scalarlike side, $c_0=m+c<0$ counterbalances the positive $a r$. 
Quantitatively, we find that in the inner region there is an important cancellation in the scalar part $S(r)$, so that the vector one is much larger. For example, 
\bea
{\rm at}\;\; r=0.5~\gev^{-1}\ : \qquad S(r)=0.025~\gev, \quad V(r)=-1.5~\gev\,.
\eea
At $r=1~$GeV$^{-1}$, the vector part is still much larger, and around $r=2~$GeV$^{-1}$ they are comparable in size. 
This implies that there is a rather extended region, say for $r\lesssim 1.5$~GeV$^{-1}$, in which the Hamiltonian is mostly vectorlike and where the approximate $\gamma_5$-symmetry is valid. 
This in fact is the inner region discussed above. This observation may suggest that the Dirac spinors themselves obey this symmetry in the same region. For sufficiently large $r$, however, the scalar potential wins and we return to a situation in which $\gamma_5$-symmetry is badly broken and both $(1/2)^-$ and $(1/2)^+$ states may retain an approximate nonrelativistic structure.

\subsection{Comparisons of integrated quantities}
We now consider the charges and radii of various distributions ${\cal O}\in\{V,A,S\}$, i.e.,
\bea\label{gr}
g_{\cal O}^{(+)} =\int d^3\vec{r}~\rho_{\cal O}^{(+)}(r)\,,  \qquad{\rm and} \qquad \langle r^2\rangle_{\cal O}^{(+)}=\int d^3\vec{r}~r^2\rho_{\cal O}^{(+)}(r) \,,
\eea 
where, again, the superscript $+$ is used when the external states are orbitally excited heavy-light mesons $j^P=(1/2)^+$ ($L=1$).
After the discussion made in the previous subsection, it is easy to anticipate as good an agreement between the integrated quantities computed on the lattice and within the two quark model frameworks considered in this paper. 
Their comparison, however, bears its own interest: (1) the integrated quantities help to summarize and quantify the comparison by a few illustrative numbers;  (2) these quantities are physically interesting, and some of them are even related to physical quantities, such as the couplings describing the strong decay to a soft pion. Radii (and higher moments) of distributions
also may have the interest of reflecting the respective extension of the distributions.

We note that such quantities are hardly sensitive to the wave function close to the origin, although that region has its specific interest as well. 
In the case of the $(1/2)^+$ state, two distributions change sign, resulting in small values for axial and scalar charges, and this is the reason why in eq.~(\ref{gr}) we do not define the radii by normalizing the moments to their respective charges, as we did in our lattice paper~\cite{emmanuel}. Furthermore, concerning the scalar density, as we discussed in a previous section, in QCD it corresponds to a logarithmically divergent quantity that should be properly renormalized. A conventionally adopted renormalization scale when discussing the physics of light quarks is $\mu=2~\gev$, that in our case coincides with the inverse lattice spacing. Therefore, as in our lattice paper~\cite{emmanuel} the lattice QCD results for $g_{S}^{(+)}$  refer to $\msbar$ renormalization scheme, and the scale $\mu=2~\gev$. That, of course, is not an issue in quark models. The results are summarized in Tab.~\ref{tab4}.

\begin{table}[h!]
\begin{center}{\scalebox{.95}{\begin{tabular}{|c||c|c|c||c|c|c|}
\cline{2-7}
\multicolumn{1}{c||}{}&\multicolumn{3}{|c||}{$\displaystyle{j^P=(1/2)^{-^{}}}$}&\multicolumn{3}{|c|}{$j^P=(1/2)^+$}\\
\hline
{\phantom{\Huge{l}}}\raisebox{-.1cm}{\phantom{\Huge{j}}}
 {\sl Quantity}   &  Dirac  & BT  & LQCD & Dirac  & BT  & LQCD\\
\hline
{\phantom{\Huge{l}}}\raisebox{-.1cm}{\phantom{\Huge{j}}}
$ \langle r^2\rangle_{V}^{(+)}$~[$\gev^{-2}$] & $4.82$  &  $5.17$ & $6.10(8)$ & $5.96$ &  $5.42$ & $4.72(42)$ \\
\hline
{\phantom{\Huge{l}}}\raisebox{-.1cm}{\phantom{\Huge{j}}}
$g_A^{(+)}$ & $0.63$ & $0.71$ & $0.58(2)$ & $\approx 0$ &$ 0.13$ & $0.08(3)$ \\
\hline
{\phantom{\Huge{l}}}\raisebox{-.1cm}{\phantom{\Huge{j}}}
$ \langle r^2\rangle_{A}^{(+)}$~[$\gev^{-2}$] & $2.72$  &  $2.87$ & $2.99(9)$ & $-1.25$ & $-0.015$ & $-0.63(37)$ \\
\hline
{\phantom{\Huge{l}}}\raisebox{-.1cm}{\phantom{\Huge{j}}}
$g_S^{(+)}$  & $0.44 $ & $0.56$ & $0.61(3)$ &$ 0.51$ & $0.30$  & $0.45(14)$ \\
\hline
{\phantom{\Huge{l}}}\raisebox{-.1cm}{\phantom{\Huge{j}}}
$ \langle r^2\rangle_{S}^{(+)}$~[$\gev^{-2}$] & $1.67$   & $1.72$  & $2.33(15)$ & $4.85$ & $2.73$ & $3.8(1.4)$ \\

\hline
\end{tabular}
}}
\caption{\footnotesize{\sl Results for moments of the densities discussed in this paper, charges and mean square radii, as obtained by using two quark model frameworks, as well as lattice QCD. Note that the vector charges are $g_V\equiv g_V^+=1$, by definition, while the scalar charges and mean square radii inferred from lattice QCD are renormalized at $\mu=2$~GeV in the $\msbar$ renormalization scheme. }}
\label{tab4}
\end{center}
\end{table} 

 \underline{Radii of the vector distributions}: For the charge distribution, the integral should be $1$ on both sides, so there is no check of the quark model by lattice QCD. As for the  positive definite mean square radii, we see that the results are rather similar for the two states.  This is in contrast to what one would expect on the basis of a nonrelativistic model, in which the centrifugal potential  [$\propto l(l+1)/r^2$] suggests that $L=1$ states should be more extended than $L=0$ ones. Once more, this emphasizes the highly relativistic character of the dynamics of the light quark. This is particularly the case with the lattice QCD data which indicate the $L=1$ states being even less extended than the $L=0$ ones.

\underline{Integral of the axial distributions}: In this case, on the contrary, the integral is physically meaningful. For the ground state $g_A$, commonly denoted 
as $\widehat g$, is an analog to $g_A$ of the nucleon axial current. $\widehat g$ has been  calculated in a number of ways. In quark models its value is known to be lower than its nonrelativistic limit, $\widehat g=1$, due to relativistic corrections in  Dirac spinors, as first stressed by P.N.~Bogolioubov~\cite{bogolioubov}. Indeed from our two frameworks, and with potentials specified above, we obtain: $\widehat g=0.63$ (Dirac) and $\widehat g=0.71$ (BT).~\footnote{For a detailed discussion related to the quark model calculation of $\widehat g$, please see ref.~\cite{dirac1999}.}
These numbers are in very good agreement with what has been obtained after integrating the radial distributions computed on the lattice (see also refs.~\cite{ghat-lattice}), and are much larger than the results inferred from light cone QCD sum rules~\cite{khodjamirianG}, but compatible with the experimentally extracted $\widehat g = 0.59(7)$~\cite{CLEO}.

As for the $(1/2)^{+}$ state, the coupling $g_A^+$ is often referred to as $\widetilde g$, and it appears to be much smaller and of the opposite sign than its nonrelativistic value, $\widetilde g=-1/3$, as  can be seen from Table~\ref{tab4}.  Such a small value is understandable because of the possible cancellation between the region in which $\rho_A^+(r)$ is positive (small $r$) and negative (large $r$). 
Let us stress that this cancellation means that the approximate ``chiral" symmetry between the $(1/2)^{+}$ and the $(1/2)^{-}$ states, which is seen for the local distributions in the inner region and for the vector radii, is completely broken in the axial case, i.e. $\widetilde g \ll \widehat g$. This cancellation is one more striking success of the quark models, already stressed in ref.~\cite{damir0406296}, and it is in contradiction with the main assumption made in  ref.~\cite{bardeen}, that  $\widetilde g=\widehat g$. 

\underline{Radii of the axial distributions}: The remarkable fact is that, within the convention adopted above~(\ref{gr}),  the axial radius of $L=1$ state is neatly negative both in quark models and in QCD. This is different from the sign of $\widetilde g$ because an extra $r^2$ factor inside the integral exacerbates the importance of the region of large $r$, in which $\rho_A^+(r)$ is negative.

\underline{Integrals and radii of the scalar distributions}:
As it has been explained above, these are not physically meaningful quantities by themselves. Only the ratios, like $\langle r^2\rangle_S/g_S$ or $g_S^+/g_S$ in which the renormalization arbitrariness is eliminated,  computed on the lattice and by using various quark models, can be compared.  In fact, on the lattice we obtained $\langle r^2\rangle_S/g_S =3.8(2)~\gev^{-2}$, that is in very good agreement with what we obtain with the Dirac model, $\langle r^2\rangle_S/g_S =3.8~\gev^{-2}$, and with the BT model, $\langle r^2\rangle_S/g_S =3.1~\gev^{-2}$. 
Similar agreement is observed for the $L=1$ state, namely the lattice value $\langle r^2\rangle^+_S/g_S^+ =8.4(3.1)~\gev^{-2}$, agrees with the quark model results, $\langle r^2\rangle^+_S/g_S^+ =9.5~\gev^{-2}$ [Dirac], $9.1~\gev^{-2}$ [BT].

We note however that other ratios are in less good agreement. For example on the lattice we obtain, $g_S^+/g_S=0.74(23)$, while the quark models give $g_S^+/g_S=1.16$ [Dirac], $0.54$ [BT]. Similarly 
$\langle r^2\rangle_S^+/\langle r^2\rangle_S =1.6(6),\ 2.9,\ 1.6$, for the lattice QCD, Dirac, and BT models, respectively. 

We note that in contrast with the axial counterpart, the integral $g_S$ for the $(1/2)^+$ state is not especially small, in spite of the fact that there is also a zero in $\rho_S^+(r)$. 
It shows that the complete cancellation found in the axial case is something nontrivial, resting on a quantitative agreement of the structure of the wave  function with what is observed on the lattice, as it can be seen from eq.~(\ref{dirac-distr-1}). With the larger coefficient of the upper components ($\propto | f_{1/2}^{1}(r)\bigr|^2$) in the scalar density  the zero appears earlier than in the axial case and therefore the small $r$ region [in which  $\rho_S^+(r)$ is negative] no longer cancels $\rho_S^+(r)$ in the large $r$ region, where it is positive. 
From the above discussion we see that the sign and magnitudes of the integrated quantities for $L=1$ states constitute a sensitive test of the internal structure of quark models because they probe the lower component of Dirac spinors.~\footnote{While making this comment we have in mind the Dirac model. The same argument also holds also for the BT model. Indeed, with a little extra work it can be shown that the BT expressions can be cast into the forms similar to the ones written in eqs.~(\ref{dirac-distr-0}, \ref{dirac-distr-1}).} 
Moreover, it can be easily understood that $\langle r^2\rangle_S^+$ is positive, since the positive contribution of large $r$ is even more favored.

\section{Conclusion} 
In this paper we made a quantitative comparison of various distributions of the light valence quark as a function of its distance from the static heavy quark within the heavy-light mesons, as  previously obtained by using lattice QCD~\cite{emmanuel}, and the corresponding quantities in two types of quark models. For this purpose we use a model based on the Dirac equation, and another one formulated in the relativistic framework of Bakamjian--Thomas. In both models the parameters in the potentials have been fixed to describe the hadronic mass spectrum only. As a result we find that the detailed comparison between the distributions for $L=0$ state [$(1/2)^-$] shows excellent agreement  of both models with the lattice counterparts. 

Moreover, the agreement also holds for the orbitally exited states, $L=1$  [$(1/2)^+$], which is highly non-trivial because the shape of the distributions are completely different, and because the expressions derived within quark models are very sensitive to the relativistic corrections and the structure of the potential. 

We focused on three bilinear light quark operators, vector, axial, and scalar, and for all of them and for both heavy-light mesons [$(1/2)^\mp$] we find the agreement to extend over a large range of distances, $r \lesssim 4~\gev^{-1}$, in which the distributions of these densities are strongly varying (they change values by $2$ orders of magnitude).

What conclusions can be drawn from this agreement?\\
\begin{itemize}
\item[--] We do not share the widespread opinion that one can get anything at will out of quark models because of an alleged large freedom. 
We repeat that our quark models were devised a long time ago, and that no adjustment has been performed on the lattice densities which are the object of the present study. 
In fact there is not much freedom in fixing the models if one wants to describe a  sufficiently large set of experimental data, for instance hadron masses. 
Moreover, in describing the light quark dynamics the inclusion of  a relativistic kinetic energy is  necessary. It has a clear impact on the structure of wave functions. Indeed, the matrix elements of current densities would not agree at all with lattice QCD if we used the nonrelativistic treatment, even if the spectra could be reasonably well fitted. In particular, one would not get zeroes for the distributions of the axial and scalar densities in the case of $L=1$ states.
In addition, the specific Lorentz structure of the potential, scalar at long-distances (confining part) and vector at short-distances, seems the simplest way to get the small spin-orbit splittings that are required for the static spectrum of heavy-light mesons. This structure is particularly important for understanding the behavior of distributions considered in this paper in the so-called inner region ($r\lesssim 1.5~\gev^{-1}$).

\item[--] Having adopted the above arguments in constructing the quark models, it may not appear too surprising that the resulting radial distributions $\rho_{\cal O}^{(+)}(r)$ (${\cal O}=V,A,S$) are similar. However it remains surprising that they agree so well and over such a large range of distances $r$. Still more surprising is their agreement with the field theoretical approach to strong interactions captured by numerical simulations of QCD on the lattice. Not only do the distributions have similar shape but they also agree quantitatively on the large range of values of $r$. In particular, their integrated characteristics, $g_{\cal O}^{(+)}$ and  $\langle r^2\rangle_{\cal O}^{(+)}$, are in good agreement with lattice QCD. 

\item[--] Finally, a comment  is in order concerning the success of the BT model used in this paper. The potential model GI, used in our BT approach, is the one formulated in ref.~\cite{godfrey}.  Its success in describing the detailed lattice data suggests that the wave functions derived from the GI model are physically adequate. Therefore, it seems likely that their use is not the source of the difficulties encountered in certain quark model calculations of heavy to light processes (inclusive rates and form factors)~\cite{bXsgamma,simula}.

\end{itemize}

We believe that the observations made in this paper further emphasize the continuing relevance of quark models for the more intuitive physical interpretation of lattice data. In addition, after testing the quark models in a detailed description of the structure of  the lowest states one can more reliably use them for understanding the experimental results which cannot yet be treated by means of lattice QCD. 
Conversely, lattice data are precious since they allow one to test models in situations which could not be accessed experimentally. The static case of QCD is a particular example where the lattice data can provide powerful tools for discrimination of various models.

\section*{Acknowledgements}
\hspace*{\parindent}
This work was supported in part by the EU Contract No. MRTN-CT-2006-035482 (Flavianet). E.C. was supported by Contract No.FIS2008-01661 from MEC (Spain) and 
FEDER.

\newpage
\section*{Appendix}
\subsection*{Dirac Model} 
In computations with the Dirac model we choose to work with the orthonormal basis of pseudo-Coulombic functions, 
\bea
\phi_n(r)=\sqrt{8 (n-1)!\over  (n+1)!} \ \beta^{3/2} \ e^{-\beta r}\  L_{n-1}^{(2)}(2\beta r)\,.
\eea
To enable the reader with sufficient information necessary to reproduce (approximately) all numerical results presented in this paper we also give the expansion coefficients in 
\bea
f_j^k(r)=\sum_{i=1}^n c_i^{(f)} \phi_i(r)\,,\quad g_j^k(r)=\sum_{i=1}^n c_i^{(g)} \phi_i(r)\,,
\eea
in the case of $n=15$, with $\beta = 0.75\ \gev$, namely,
\bea
j^P=(1/2)^-:&& \nn\\
c_i^{(f)} &\in& \{0.842464, 0.0518048, -0.0876299, -0.0303635, 0.0110048,\nn \\
&&0.0154663, 0.00653305, -0.000644675, -0.00299685, -0.00252853, \nn\\
&&-0.00156505, -0.00125756, -0.00165585, -0.00213105, -0.00180187\}\,,\nn\\
c_i^{(g)} &\in& \{-0.514038, 0.0450279, 0.103569, 0.0373234, -0.000273779, \nn\\
&&-0.00435446, 0.002004, 0.00634121, 0.00680183, 0.00527419, \nn\\
&&0.00342261, 0.00181708, 0.000444552, -0.000628798, -0.00097132\}\,\\
j^P=(1/2)^+&:& \nn\\
c_i^{(f)}& \in& \{0.82558, -0.196543, -0.184425, -0.022037, 0.0312719, \nn \\
&&0.0163965, -0.00538404, -0.0133003, -0.0106817, -0.00538832, \nn \\
&&-0.0020597, -0.00168964, -0.00317236, -0.00454244, -0.00383235\}\, \nn \\
c_i^{(g)} & \in& \{0.376853, 0.307429, 0.061718, -0.0414061, -0.0369034, \nn\\
&&-0.00895755, 0.00599181, 0.00698936, 0.00275505, -0.000804702, \nn\\
&&-0.00233478, -0.00283731, -0.00339297, -0.00398404, -0.00340649\}\ .\nn
\eea
We stress again that the numerical results presented in this paper are however obtained in the much larger and precise basis ($n\geq 40$).
\subsection*{BT Model} 
In computations with the BT model, for the ground states, we computed the ground state wave function [$j^P=(1/2)^-$] by expanding in a truncated harmonic oscillator basis,
\bea
\label{16e}
\varphi(\vec{k}) = \sum_{i=0}^{n=15} C_i (-1)^i (4\pi)^{3/4}2^i\ \sqrt{{(i!)^2 \over (2i+1)!}} {1 \over \beta^{3/2}} L{_i^{1/2}}\left({\vec{k}^2 \over {2 \beta^2}}\right) exp\left(-{\vec{k}^2 \over {2 \beta^2}}\right)\,,
\eea
\\
With the masses $m_2 = 0.419~\gev$ for the strange quark, and in the heavy quark limit ($m_1 = 10^4~ \gev$), we find the following values for the coefficients:
\bea
\label{15e}
C_i & \in& \{0.961012542, 0.199752584, 0.158164397, 7.29950815\times 10^{-2}, \nn \\  
&& 5.70315719\times 10^{-2},  3.4423735\times 10^{-2}, 2.76433732\times 10^{-2}, \nn \\
&& 1.94679499\times 10^{-2}, 1.55471498\times 10^{-2}, 1.18442774\times 10^{-2}, \nn \\
&& 9.54456627\times 10^{-3}, 7.95029942\times 10^{-3}, 5.91434911\times 10^{-3}, \nn \\
&& 5.39294863\times 10^{-3}, 3.7691386\times 10^{-3}, 3.7654086\times 10^{-3}\}\,.\nn
\eea
Concerning the excited states, $j^P=(1/2)^+$, we expand in the basis:
\bea
\label{26e}
\varphi_{1/2}(\vec{k}) = \sum_{i=0}^{15} D_i (-1)^i (4\pi)^{3/4}2^{i+1}\ \sqrt{{i!(i+1)! \over (2i+3)!}} {|\vec{k}| \over \beta^{5/2}}\ L{_i^{3/2}}\left({\vec{k}^2 \over {2 \beta^2}}\right) exp\left(-{\vec{k}^2 \over {2 \beta^2}}\right)\,,
\eea
and find the coefficients to be 
\bea
\label{27e}
D_i & \in& \{  0.9684052, 0.1689021, 0.1572283, 6.3688330 \times 10^{-2}, \nn \\
&&5.2452430 \times 10^{-2},  2.9747754 \times 10^{-2}, 2.3702351 \times 10^{-2}, \nn \\
&&1.6382594 \times 10^{-2}, 1.2990386 \times 10^{-2}, 9.7295996 \times 10^{-3},  \nn \\
&&7.6334681 \times 10^{-3}, 6.3195196 \times 10^{-3}, 4.7800541 \times 10^{-3}, \nn \\
&&4.1553467 \times 10^{-3}, 3.0716809 \times 10^{-3}, 2.9703260 \times 10^{-3}\}.
\eea

\end{document}